\documentclass[11pt]{amsart}

\usepackage[margin=2.8cm]{geometry}
\usepackage{amsmath,amssymb,amsthm,mathtools}
\usepackage{enumitem}
\usepackage{graphicx}
\usepackage{booktabs}
\usepackage[colorlinks=true,linkcolor=blue,citecolor=blue,urlcolor=blue]{hyperref}
\usepackage[section]{placeins}
\usepackage[capitalise]{cleveref}
\usepackage{algorithm}
\usepackage{algpseudocode}
\theoremstyle{plain}
\newtheorem{theorem}{Theorem}[section]
\newtheorem{proposition}[theorem]{Proposition}

\newtheorem{corollary}[theorem]{Corollary}
\theoremstyle{definition}
\newtheorem{definition}[theorem]{Definition}

\newtheorem{remark}[theorem]{Remark}
\numberwithin{equation}{section}

\newcommand{\R}{\mathbb{R}}
\newcommand{\sym}{\mathrm{sym}}

\DeclareMathOperator{\Cov}{Cov}
\DeclareMathOperator{\Var}{Var}

\newcommand{\E}{\mathbb{E}}

\begin{document}

\title[Cotangent-lift inference of stochastic microbial dynamics]{Cotangent-lift inference of stochastic microbial dynamics}

\author{José A. Capit\'an}
\address{Department of Applied Mathematics and Complex Systems Group, Universidad Polit\'ecnica de Madrid, Avenida Juan de Herrera 6, 28040 Madrid, Spain}
\email{ja.capitan@upm.es}

\author{Alberto Meg\'ias}
\address{Department of Applied Mathematics and Complex Systems Group, Universidad Polit\'ecnica de Madrid, Avenida Juan de Herrera 6, 28040 Madrid, Spain}
\email{alberto.megias@upm.es}
\thanks{Corresponding author: Alberto Meg\'ias (\texttt{alberto.megias@upm.es}).}

\author{Cristina Sard\'on}
\address{Department of Applied Mathematics, Universidad Polit\'ecnica de Madrid, Avenida Juan de Herrera 6, 28040 Madrid, Spain}
\email{mariacristina.sardon@upm.es}
\date{\today}

\begin{abstract}
We study the parameter estimation for the stochastic logistic and Lotka--Volterra diffusions from discrete, irregularly spaced observations corrupted by measurement noise, a setting typical of microbiome time series. For the logistic equation, the growth rate acts as a time change and does not enter the stationary law, so it can be identified only from the temporal structure of the data. We prove that the maximum-likelihood estimator of the Euler scheme, with the observations taken as the states, coincides with the regression of log-increments on abundances used in the literature, and we characterise what this estimator cannot see under measurement noise: for the linearised dynamics, the law of consecutive pairs of observations is invariant along an explicit one-parameter family of relaxation rates and noise levels, and the regression converges to a limit set by the noise. Treating the true abundances as latent, we compute the likelihood with a Gaussian moment filter and show that measurement noise is precisely what couples the filter states across samples. The exact gradient is then the backward recursion of the discrete cotangent lift, whose cost, two to three likelihood evaluations, does not depend on the number of parameters, and we give the covector map of each filter step in closed form. In simulations, the latent estimator recovers the relaxation times, noise levels and interaction strengths that the regression biases, while larger communities require prior information on the interactions. In gut microbiome time series it attributes most of the variability of a dominant genus to measurement noise and estimates its relaxation time at about $64$ hours, with a large standard error.
\end{abstract}

\maketitle


\section{Introduction}
\label{sec:intro}

Longitudinal microbiome data are among the densest ecological time series available. Daily samples, sequenced over several months, give hundreds of time points for each of dozens of taxa \cite{Caporaso2011}. Their statistics show robust regularities: the temporal fluctuations of the abundance of a taxon follow a Gamma distribution, their variance follows Taylor's law \cite{Grilli2020,Ji2020} with an exponent close to two, and mean abundances across taxa are approximately lognormal \cite{Grilli2020}. Remarkably, these regularities are the stationary statistics of a model without interactions, a stochastic logistic equation for each taxon driven by multiplicative environmental noise \cite{Grilli2020,Descheemaeker2020}. When interactions are included, the natural extension is the stochastic Lotka--Volterra (LV) system. Its long-time behaviour under environmental noise is well understood, with sharp criteria for persistence and extinction \cite{HeningNguyen2018}. In the limit of many species with random interactions, LV dynamics has been analysed with the methods of statistical physics. Dynamical mean-field theory, of which \cite{Galla2024} gives a step-by-step derivation, reduces the community to an effective equation for one representative species and yields a phase diagram with a unique stable equilibrium, multiple attractors \cite{Bunin2017,Biroli2018,MegiasEtAl2026}, or unbounded growth \cite{Galla2024}. At every level of description, the parameters of these models (growth rates, relaxation times, interaction coefficients) have to be inferred from time series. This inverse problem is the subject of the present paper.

The usual method is linear regression. To perform a linear regression on this dynamics, one can write $\zeta_i=\log X_i$ for the log-abundance of species $i$. Then, over a sampling interval $h$, the Euler discretisation of the LV equations approximates the log-increment $(\zeta_i(t+h)-\zeta_i(t))/h$ by $r_i+\sum_jA_{ij}X_j(t)$ plus noise. This is linear in the unknown growth rates $r_i$ and interactions $A_{ij}$, which are then estimated by (regularised) least squares \cite{Stein2013,FisherMehta2014,Bucci2016}. The structural identifiability of the generalised LV model in this setting is studied in \cite{Remien2021}, and multivariate autoregressive models for log-abundances, widely used in ecology, rest on the same idea \cite{Ives2003}. From the statistical point of view, the regression is the Euler quasi-likelihood estimator of a discretely observed diffusion, whose consistency and discretisation bias are classical \cite{FlorensZmirou1989,Kessler1997,BibbySorensen1995}. That theory assumes, however, that the diffusion on the log-abundances is observed exactly.

However, in microbiome data, there are sources of noise that makes this measurements to be inexact. Every sample goes through DNA extraction, amplification and sequencing at finite depth, and each of these steps adds variability that has nothing to do with the dynamics. Moreover, samples are taken about once a day, with gaps of several days, while the relaxation times of interest are themselves of the order of a day. We therefore distinguish the true log-abundance $\zeta(t)$, defined at every time but never observed, from the measurements $Y_k=\zeta(t_k)+\varepsilon_k$. The trajectory $\zeta$ is a \emph{latent} variable. We call an estimator \emph{observed-state} if it takes $Y_k$ as the state of the system, and \emph{latent-state} if it treats $\zeta$ as unknown and accesses it only through the measurements. Diffusions observed with additive errors have been studied in mathematical statistics, where the measurement error changes the asymptotic structure of the statistical experiment and the attainable rates of convergence \cite{GloterJacod2001}, and in financial econometrics, where microstructure noise dominates high-frequency increments and makes estimators based on them inconsistent \cite{AitSahalia2005}. In population ecology, observation error is known to bias estimates of density dependence, to the point of producing spurious density dependence in census data \cite{KnapeDeValpine2012}, and to be difficult to separate from process noise. This motivated state-space models in which the population trajectory is latent \cite{deValpineHastings2002,Dennis2006,AugerMethe2021}. In microbiome research, MDSINE2 \cite{Gibson2025} is a fully Bayesian model of this kind, with latent abundance trajectories, a negative-binomial model for read counts, and qPCR measurements of total biomass.

A latent-state likelihood is computed by a filter, which propagates a belief about the hidden state, a mean $m$ and a covariance $P$, between samples and corrects it at each measurement. Gaps, missing values and known perturbations enter this computation naturally. Because the measurements are noisy, the filter carries memory from one sample to the next, so the likelihood depends on the parameters through a long chain of computed states. A community of $S$ species has of the order of $S^2$ parameters, and a gradient computed by finite differences, or by propagating sensitivities forwards, costs of the order of $S^2$ evaluations of the likelihood. The alternative is to propagate sensitivities backwards. For a computation $Z_{i+1}=\Phi_i(Z_i,u)$, the adjoint of the Jacobian of each step maps covectors at its output to covectors at its input. By analogy with the continuous adjoint system, which Tran and Leok describe as the cotangent lift of a vector field \cite{TranLeok2024}, we call this map the \emph{discrete cotangent lift}; its iteration from the end of the computation to its beginning is the discrete adjoint method. The same construction underlies variational data assimilation in meteorology \cite{LeDimetTalagrand1986}, coincides with reverse-mode automatic differentiation \cite{GriewankWalther2008}, and preserves the canonical pairing between variations and covectors, which links it with symplectic integration \cite{SanzSerna2016}. For linear Gaussian state-space models the score can be computed with one forward and one backward pass \cite{SegalWeinstein1989}, general state-space likelihoods are routinely differentiated by automatic differentiation \cite{Kristensen2016}, and adjoint methods for pathwise gradients of stochastic differential equations are developed in \cite{LiWongChenDuvenaud2020}.

\subsection*{Main results}
We summarise our contributions and their relation to the literature.

\begin{enumerate}[label=(\roman*), leftmargin=2.2em]
\item \emph{Exact gradients by the cotangent lift} (\cref{sec:lift}). The backward recursion of \cref{thm:lift} gives the exact gradient of a cost accumulated along a computed trajectory; when the states are data, it collapses to a sum of one-step terms (\cref{cor:observed}). This is the discrete adjoint method \cite{GriewankWalther2008,SanzSerna2016,TranLeok2024}; the dichotomy between the two cases organises the paper.

\item \emph{The growth rate is invisible at stationarity} (\cref{sec:models}). For the stochastic logistic equation the growth rate acts as a time change (\cref{prop:clock}), and the Gamma stationary law, which exists if and only if $\sigma<1$ (\cref{thm:gamma}), does not involve it. The growth rate is therefore identifiable from the law of samples at a positive lag or from a transient, but not from the stationary marginal (\cref{thm:ident}). This is the single-species counterpart of the stationary blindness of interactions in \cite{CapitanMegiasSardon2026}.

\item \emph{The regression is the observed-state lift} (\cref{sec:regression}). The maximum-likelihood estimator of the Euler scheme with the observations taken as states is the weighted regression of log-increments on abundances (\cref{thm:ols}) used in the microbiome literature \cite{Stein2013,FisherMehta2014,Bucci2016}. Its asymptotic covariance and biases, instances of the classical theory \cite{FlorensZmirou1989,Kessler1997,MarriottPope1954,TangChen2009}, are given in Section~S1 of the Supplementary Material.

\item \emph{What pairs of noisy samples cannot identify} (\cref{prop:noise}). For the linearised dynamics observed with additive noise at interval $h$, the law of consecutive samples is invariant along an explicit one-parameter family of relaxation rates, variances and noise levels, the regression converges to a limit fixed by the lag-$h$ autocorrelation of the measurements, and two lags restore identifiability. The bias of density dependence under observation error is known in ecology \cite{Dennis2006,KnapeDeValpine2012,Ives2003}, and measurement errors in diffusions are analysed asymptotically in \cite{GloterJacod2001}; the proposition makes the phenomenon exact at finite $h$.

\item \emph{Measurement noise makes the backward sweep necessary} (\cref{sec:latent}). For a latent-state likelihood computed by a Gaussian moment filter, the sensitivity of each corrected state to the predicted one vanishes as the measurement noise tends to zero, in which limit the latent likelihood reduces to the observed-state one (\cref{prop:memory}). Measurement noise thus creates the chain of dependencies, and the exact gradient is one backward sweep, at a cost independent of the number of parameters (\cref{thm:latent}), with the covector maps given in closed form in \cref{app:vjp}. It costs two to three likelihood evaluations, agrees with automatic differentiation \cite{Kristensen2016} to $4\times10^{-15}$, and, unlike the linear Gaussian case \cite{SegalWeinstein1989}, applies to a filter that is nonlinear in the state and in the parameters.

\item \emph{Numerical and empirical evidence} (\cref{sec:synthetic,sec:data}). In simulations the regression converges to the limits predicted by \cref{prop:noise}, whereas the latent estimator recovers relaxation times, noise levels and, for ten species, unshrunk interactions. In the gut series of \cite{Caporaso2011}, measurement noise is significant for six of twelve genera, and for the dominant genus of one subject the relaxation time is $64\pm34$ hours instead of the one-step value of $19$ hours \cite{Grilli2020}. Unlike MDSINE2 \cite{Gibson2025}, the approach uses maximum likelihood and requires neither total-biomass data nor a count model, at the price of a simpler measurement model.
\end{enumerate}

We wish to be explicit about what is standard. The cotangent lift is reverse-mode differentiation, the latent-state formulation is a state-space model in the sense of population ecology \cite{deValpineHastings2002,AugerMethe2021}, and the filter is an assumed-density Gaussian filter \cite{Sarkka2013}. What is new is the identification of the microbiome regression with the observed-state lift (iii), the exact finite-$h$ description of what pairs of noisy samples identify (iv), the proof that measurement noise is precisely what creates memory in the filter and makes the backward sweep necessary (v), and the closed-form backward maps for the stochastic LV moment filter, with the treatment of the symmetrisation that both passes require.

The paper is organised as follows. \Cref{sec:lift} introduces the discrete cotangent lift and proves that it yields the exact gradient. \Cref{sec:models} recalls the stochastic logistic and LV models and establishes their identifiability properties. \Cref{sec:regression} shows that the observed-state lift is the regression and analyses its failure under measurement noise. \Cref{sec:latent} constructs the latent-state likelihood and its gradient. \Cref{sec:synthetic,sec:data} test the method on simulated communities and on the gut series of \cite{Caporaso2011}, and \cref{sec:discussion,sec:concl} discuss its scope and limitations.

\section{The discrete cotangent lift}
\label{sec:lift}

Suppose a quantity is computed in many small steps, and we want to know how the final result depends on a set of parameters. A simple example is a population projected forward day by day with a growth model. The answer is the chain rule, but how the chain rule is organised makes an enormous difference when there are many parameters. This section explains the organisation we use.

Let the state of the computation be $Z$, which may be a vector, a matrix or a pair of both, and let $u\in\R^p$ collect the parameters. Each step produces a new state from the previous one,
\[
  Z_{i+1}=\Phi_i(Z_i,u),\qquad i=0,\ldots,n-1,
\]
possibly using fixed data such as a measurement, which we do not write. Along the way a cost is accumulated, for instance a sum of prediction errors, and we want its gradient with respect to $u$.

Consider a single step and ask how a small change propagates through it. If the input changes by $\delta Z$ and the parameters by $\delta u$, then, to first order, the output changes by $\delta Z^{+}=D_Z\Phi\,\delta Z+D_u\Phi\,\delta u$. Now suppose we already know how sensitive the cost is to the output of this step. We encode this as a covector $\lambda^{+}$, so that the cost changes by $\langle \lambda^{+},\delta Z^{+}\rangle$.\footnote{Throughout the paper a superscript $+$ marks the value \emph{after} a step, and the same symbol without it the value before.} Here $\langle\cdot,\cdot\rangle$ is an inner product; concretely, we will use explicitly the usual vector dot product when the state is a vector, and the trace pairing $\langle\Lambda,\delta P\rangle=\mathrm{tr}(\Lambda^\top\delta P)$ when it is a matrix, but it can be generalized. Substituting the expression for $\delta Z^{+}$ splits this change in two:
\begin{equation}
  \langle \lambda^{+},\delta Z^{+}\rangle
  =
  \langle D_Z\Phi^{*}\lambda^{+},\delta Z\rangle
  +
  \langle D_u\Phi^{*}\lambda^{+},\delta u\rangle .
  \label{eq:pull}
\end{equation}
The first term says how sensitive the cost is to the \emph{input} of the step: the covector has been carried one step back, by the transpose $D_Z\Phi^*$ of the step's Jacobian. The second term is the step's direct contribution to the gradient. Carrying covectors backwards in this way is the discrete counterpart of what Tran and Leok call the cotangent lift \cite{TranLeok2024}. The covectors $\lambda_i$ are the discrete adjoint variables, and the backward recursion is the discrete adjoint equation of the computation. It is also exactly what reverse-mode automatic differentiation does \cite{SanzSerna2016,GriewankWalther2008}. Repeating the operation step after step, from the end of the computation to its beginning, gives the full gradient. We record the two notions just used.

\begin{definition}[Covector and cotangent lift]
\label{def:lift}
Let $\mathcal Z$ be a finite-dimensional real vector space with an inner product $\langle\cdot,\cdot\rangle$. A \emph{covector} at a state $Z\in\mathcal Z$ is a linear functional on variations $\delta Z$. We represent it, through the inner product, by an element $\lambda\in\mathcal Z$ acting as $\delta Z\mapsto\langle\lambda,\delta Z\rangle$. For a $C^1$ step $\Phi:\mathcal Z\times\R^p\to\mathcal Z$, the \emph{cotangent lift} of $\Phi$ at $(Z,u)$ is the linear map
\[
  \lambda^{+}\longmapsto\bigl(D_Z\Phi(Z,u)^{*}\lambda^{+},\;D_u\Phi(Z,u)^{*}\lambda^{+}\bigr)\in\mathcal Z\times\R^p ,
\]
which sends a covector at the output to a covector at the input, together with a contribution to the gradient with respect to the parameters. It is characterised by \eqref{eq:pull}.
\end{definition}

\begin{theorem}[Exact gradient along a computed trajectory]
\label{thm:lift}
Let $Z_0$ be independent of $u$, let $Z_{i+1}=\Phi_i(Z_i,u)$ for $i=0,\ldots,n-1$, and let
\[
  J(u)=\sum_{i=0}^{n-1}\ell_i(Z_i,Z_{i+1},u),
\]
with $\Phi_i$ and $\ell_i$ continuously differentiable. Starting from $\lambda_n=0$ and $g_n=0$, define backwards for $i=n-1,\ldots,0$
\begin{align}
  \lambda_i^{+}
  &=
  \lambda_{i+1}+\partial_{Z_{i+1}}\ell_i,
  \label{eq:precov}\\
  g_i
  &=
  g_{i+1}
  +
  D_u\Phi_i^{*}\lambda_i^{+}
  +
  \partial_u\ell_i,
  \label{eq:eta}\\
  \lambda_i
  &=
  D_{Z_i}\Phi_i^{*}\lambda_i^{+}
  +
  \partial_{Z_i}\ell_i.
  \label{eq:prec}
\end{align}
Then $g_0=\nabla_uJ(u)$. More generally, regard the part of the cost that is still to come, $J_i=\sum_{k\ge i}\ell_k$, as a function of the current state $Z_i$ and of $u$. Then $\lambda_i$ is its gradient with respect to $Z_i$, and $g_i$ its gradient with respect to $u$ at fixed $Z_i$.
\end{theorem}

The three lines have a simple reading. $\lambda_i^+$ is the sensitivity of the remaining cost to the state just after step $i$. $\lambda_i$ is the same sensitivity just before step $i$. And $g_i$ collects the contributions to the gradient from all steps from $i$ onwards.

\begin{proof}
The idea is to prove the last statement, which describes what $\lambda_i$ and $g_i$ mean, and to obtain the gradient as its special case $i=0$. Fix $i$ and regard the remaining cost $J_i=\sum_{k\ge i}\ell_k$ as a function of the current state $Z_i$ and of $u$: once $Z_i$ is given, the recursion determines everything that follows. We claim that, for every small change $(\delta Z_i,\delta u)$,
\begin{equation}
  \delta J_i=\langle \lambda_i,\delta Z_i\rangle+g_i\cdot\delta u .
  \label{eq:induction}
\end{equation}
At the end of the computation, $i=n$, there is no cost left and $\lambda_n$ and $g_n$ vanish, so the claim holds trivially. Suppose it holds at $i+1$. Since $J_i$ is the current term plus the remaining cost, $J_i=\ell_i(Z_i,Z_{i+1},u)+J_{i+1}$, the chain rule and the claim at $i+1$ give
\[
  \delta J_i
  =
  \langle\partial_{Z_i}\ell_i,\delta Z_i\rangle
  +
  \langle\partial_{Z_{i+1}}\ell_i+\lambda_{i+1},\delta Z_{i+1}\rangle
  +
  \bigl(\partial_u\ell_i+g_{i+1}\bigr)\cdot\delta u .
\]
The covector in the middle term is $\lambda_i^+$. But $Z_{i+1}$ is not free: it is produced by the step, so $\delta Z_{i+1}=D_{Z_i}\Phi_i\,\delta Z_i+D_u\Phi_i\,\delta u$. Using \eqref{eq:pull} to rewrite the middle term and collecting what multiplies $\delta Z_i$ and $\delta u$, we obtain exactly $\lambda_i$ and $g_i$ as defined in \eqref{eq:eta}--\eqref{eq:prec}. This proves the claim at $i$, and by induction for all $i$. At $i=0$ the remaining cost is the whole cost, and $\delta Z_0=0$ because the initial state does not depend on $u$. Hence $\delta J=g_0\cdot\delta u$, i.e.\ $g_0$ is the gradient.
\end{proof}

It is worth comparing this with the obvious alternatives, because the comparison is the reason for using the lift at all. The simplest is to perturb each of the $p$ parameters in turn and rerun the computation (finite differences). This costs $2p$ runs and is only approximately accurate. A more refined option is to carry the derivatives of the state with respect to every parameter forward alongside the computation (forward sensitivities). That is exact, but it means propagating a matrix with $p$ columns, so the cost again grows in proportion to $p$. The backward sweep of \cref{thm:lift} propagates a single covector, the size of one state, whatever the number of parameters, and in practice it costs a small multiple of one forward run \cite{GriewankWalther2008}. When the parameters include an interaction matrix---as the applications will show---so that $p$ grows like the square of the number of species, only the backward sweep remains practical.

There is one situation in which the sweep has nothing to do: when the states entering the cost are themselves data. Then they do not depend on the parameters, no sensitivity is passed from one step to the next, and the gradient is just the sum of what each term contributes directly. This is summarized in the following Corollary.

\begin{corollary}[Observed states]
\label{cor:observed}
If the states $Z_0,\ldots,Z_n$ are data and each $\ell_i$ depends on $u$ only through a one-step prediction $\mathcal Q_i(Z_i,u)$ of $Z_{i+1}$ from $Z_i$ (for instance its predicted mean and variance), then
\[
  \nabla_uJ(u)=\sum_{i=0}^{n-1}D_u\mathcal Q_i(Z_i,u)^{*}\,\nabla_{\mathcal Q}\ell_i .
\]
\end{corollary}

\begin{proof}
Formally, this is \cref{thm:lift} with steps $\Phi_i(Z_i,u)\equiv Z_{i+1}$ that return the next datum, so that $D_{Z}\Phi_i=0$ and $D_u\Phi_i=0$. Every term that transmits sensitivity from one step to the previous one then vanishes, and \eqref{eq:eta} reduces to $g_i=g_{i+1}+\partial_u\ell_i$. Hence $\nabla_uJ=\sum_i\partial_u\ell_i$, and the chain rule through $\mathcal Q_i$ gives the formula.
\end{proof}

This corollary is the situation of the regression methods discussed in the introduction. For the models of this paper the resulting equations even have a closed-form solution (\cref{thm:ols}). The backward sweep becomes essential only when the cost involves states that are computed rather than measured. The case we care about is a hidden trajectory reconstructed from noisy and sparse measurements, which is the subject of \cref{sec:latent}. 

\section{The models}
\label{sec:models}

\subsection*{The stochastic logistic equation.} The single-taxon model of \cite{Grilli2020} describes the abundance $X_t$ of a taxon with growth rate $r=1/\tau$, carrying capacity $K$ and a dimensionless noise parameter $\sigma>0$:
\begin{equation}
  \mathrm{d}X_t
  =
  r X_t\Bigl(1-\frac{X_t}{K}\Bigr)\,\mathrm{d}t
  +
  \sqrt{2r\sigma^2}\, X_t\,\mathrm{d}W_t.
  \label{eq:logistic}
\end{equation}
The noise is multiplicative, as befits environmental fluctuations that act on per-capita rates. It becomes additive if we work with the logarithm $Z=\log X$, the Lamperti transformation~\cite{Lamperti1964}. By It\^{o}'s formula,
\begin{equation}
  \mathrm{d}Z_t=\bigl(a-b\,e^{Z_t}\bigr)\,\mathrm{d}t+\sqrt{s}\,\mathrm{d}W_t,\qquad
  a=r(1-\sigma^2),\quad b=r/K,\quad s=2r\sigma^2 .
  \label{eq:lamperti}
\end{equation}
The ecologically relevant regime is $\sigma<1$, in which the taxon persists (\cref{thm:gamma}), and we assume it from now on, except in \cref{thm:gamma}, which covers every $\sigma>0$. The parameters $(a,b,s)$ are then just a convenient relabelling of $(r,K,\sigma)$. Every $(r,K,\sigma)$ with $r,K>0$ and $\sigma\in(0,1)$ corresponds to exactly one positive triple $(a,b,s)$, and conversely
\begin{equation}
  r=a+\tfrac12 s,\qquad \sigma^2=\frac{s}{2a+s},\qquad K=r/b .
  \label{eq:inverse}
\end{equation}
We call $\tau=1/r$ the \emph{relaxation time} of the taxon; it is the relaxation time of the deterministic logistic equation near $K$. With noise, the fluctuations of $\log X$ about the stationary state relax at the slower rate $\kappa=r(1-\sigma^2)=(1-\sigma^2)/\tau$ (\cref{sec:regression}). This model will be treated as a null model for non-interacting species.

\subsection*{The Lotka--Volterra diffusion.} For $S$ interacting taxa with abundances $X_i$, growth rates $r_i$, interaction matrix $A$ and independent environmental noises of intensity $\sigma_i$,
\begin{equation}
  \mathrm{d}X_i
  =
  X_i\Bigl(r_i+\sum_{j=1}^{S}A_{ij}X_j\Bigr)\,\mathrm{d}t
  +
  \sigma_i X_i\,\mathrm{d}W_i,
  \qquad\text{i.e.}\qquad
  \mathrm{d}\zeta_i=\Bigl(c_i+\sum_jA_{ij}e^{\zeta_j}\Bigr)\mathrm{d}t+\sigma_i\,\mathrm{d}W_i
  \label{eq:glv}
\end{equation}
in the log-coordinates $\zeta=\log X$, with $c_i=r_i-\sigma_i^2/2$. We treat $A$ as fixed. If it were drawn from a random ensemble, as in mean-field theory, what we estimate is the particular draw. Note that $\sigma_i$ here is the noise intensity of the log-abundance, whereas the $\sigma$ of \eqref{eq:logistic} is dimensionless. The logistic equation is the single-species case, with $c=a$, $A_{11}=-b$ and $\sigma_1^2=s=2r\sigma^2$.

Before turning to estimation, it helps to understand how the logistic model encodes its three parameters, because this tells us what kind of data can reveal each of them. The first observation is that the growth rate acts as a clock.

\begin{proposition}[The growth rate is a clock]
\label{prop:clock}
Let $X^{(r)}$ denote the solution of \eqref{eq:logistic} with parameters $(r,K,\sigma)$ and a given initial law. Then $(X^{(r)}_t)_{t\ge0}$ has the same law as $(X^{(1)}_{rt})_{t\ge0}$.
\end{proposition}

In other words, changing $r$ does not change what the fluctuations look like, only how fast they unfold. A taxon with $r=2$ per day behaves exactly like one with $r=1$ filmed at double speed.

\begin{proof}
Run the $r=1$ process $r$ times faster, $Y_t=X^{(1)}_{rt}$. In a short time $\mathrm dt$, $Y$ advances the original process by $r\,\mathrm dt$. Its drift is therefore multiplied by $r$, and its noise increment $\mathrm dW_{rt}$ has variance $r\,\mathrm dt$, so it can be written as $\sqrt r\,\mathrm dB_t$ with a standard Wiener process $B$. This gives $\mathrm dY_t=rY_t(1-Y_t/K)\,\mathrm dt+\sqrt{2r\sigma^2}\,Y_t\,\mathrm dB_t$, which is \eqref{eq:logistic} with growth rate $r$. Since the coefficients are locally Lipschitz and the solution neither explodes nor reaches zero (\cref{thm:gamma}), solutions are unique, and $Y$ and $X^{(r)}$ have the same law.
\end{proof}

The second observation concerns the long-run behaviour, which explains the Gamma-distributed fluctuations seen in the data.

\begin{theorem}[Invariant Gamma law]
\label{thm:gamma}
The solution of \eqref{eq:logistic} neither explodes nor reaches zero in finite time. If $\sigma^2<1$ it fluctuates forever around a stable level (it is positive recurrent), and its long-run distribution is the Gamma law with shape $\alpha=\sigma^{-2}-1$ and rate $\beta=(K\sigma^2)^{-1}$. This law has mean $K(1-\sigma^2)$ and second moment $K^2(1-\sigma^2)$, and it does not depend on $r$. If $\sigma^2=1$ the process is null recurrent, and if $\sigma^2>1$ it goes extinct asymptotically, $X_t\to0$ almost surely.
\end{theorem}

\begin{proof}
For a one-dimensional diffusion $\mathrm dX=\mu(X)\mathrm dt+g(X)\mathrm dW$, the long-run behaviour is governed by two functions \cite[Ch.~5.5]{KaratzasShreve1991}: the scale density $\mathfrak s'(x)=\exp(-\int^x2\mu/g^2)$, which tells whether the process can drift off to a boundary, and the speed density $\mathfrak m(x)=2/(g^2\mathfrak s')$, which, when normalisable, is the stationary distribution. For \eqref{eq:logistic}, $2\mu/g^2=\sigma^{-2}(x^{-1}-K^{-1})$, so
\begin{equation}\label{eq:proof_gamma}
  \mathfrak s'(x)=x^{-1/\sigma^2}e^{x/(K\sigma^2)},\qquad \mathfrak m(x)=\frac{1}{r\sigma^2}\,x^{1/\sigma^2-2}e^{-x/(K\sigma^2)} .
\end{equation}
That neither boundary is reached in finite time is easiest to see in the log coordinate \eqref{eq:lamperti}. There the noise is additive, the drift $a-be^z$ tends to $-\infty$ as $z\to+\infty$, and it is bounded and Lipschitz on every half-line $(-\infty,z_0]$. A diffusion with these properties neither explodes to $+\infty$ nor reaches $-\infty$ in finite time, so $X=e^Z$ stays in $(0,\infty)$ at all times. The integral of $\mathfrak s'$ diverges at infinity in all cases, so the process does not escape upwards in the long run either. At zero it diverges exactly when $\sigma^2\le1$, in which case the process is recurrent. For $\sigma^2>1$ it converges, and the process is then attracted to zero, $X_t\to0$ almost surely \cite[Prop.~5.5.22]{KaratzasShreve1991}. When $\sigma^2<1$ the speed density is integrable (its exponent at zero, $1/\sigma^2-2$, exceeds $-1$), the process is positive recurrent, and normalising $\mathfrak m$ gives the Gamma density. The prefactor $1/(r\sigma^2)$ cancels in the normalisation. When $\sigma^2=1$ the speed density behaves like $x^{-1}$ near zero, is not integrable, and the process is null recurrent.
\end{proof}

Note that \eqref{eq:proof_gamma} is no more than the zero-flux steady state solution of the Fokker--Planck equation associated to the stochastic logistic process \eqref{eq:logistic}.

Together, the last two results say something quite strong about what the data can reveal. We use the word identifiability in its standard sense, that is, related to an injective map.

\begin{definition}[Identifiability]
\label{def:ident}
Let $\{\mathbb P_u:u\in\Theta\}$ be the family of laws of some observable quantity, for instance the stationary marginal, or the joint law of two samples a time $h$ apart. A function $q(u)$ of the parameters is \emph{identifiable} from this family if $\mathbb P_u=\mathbb P_{u'}$ implies $q(u)=q(u')$. The whole parameter is identifiable if this holds for $q(u)=u$, i.e.\ if $u\mapsto\mathbb P_u$ is injective.
\end{definition}

In this language, the two results above already tell us a good deal. A histogram of the abundances, however long the record, identifies $K$ and $\sigma$ but contains no information at all about $r$. The growth rate can only be learned from the temporal order of the observations, from how quickly the population forgets where it was.

\begin{theorem}[Identifiability from exact observations]
\label{thm:ident}
(i) The stationary law determines $(K,\sigma)$, and the Fisher information it carries about $r$ is zero. (ii) For any fixed lag $h>0$, the joint law of the stationary pair $(X_0,X_h)$ determines $(r,K,\sigma)$. (iii) For fixed $(K,\sigma)$ and a fixed initial law with finite second moment whose mean is not constant in time, the mean curve $t\mapsto\E[X_t]$ determines $r$.
\end{theorem}

\begin{proof}
For (i), the shape of the Gamma law fixes $\sigma^2$ and its rate then fixes $K$. Since the law does not involve $r$, its score with respect to $r$ vanishes. For (ii), the distribution of $X_0$ gives $(K,\sigma)$ as in (i). By \cref{prop:clock}, observing the process at lag $h$ with rate $r$ is the same as observing the $r=1$ process at lag $rh$, and the law of the $r=1$ process depends only on $(K,\sigma)$. Let $\pi$ be the Gamma law of \cref{thm:gamma}, and let $X$ be the $r=1$ process started from $X_0\sim\pi$, which is then stationary. Ergodicity is a property of a stationary process relative to its invariant measure. The process $X$ is \emph{ergodic with respect to $\pi$} if every event that is invariant under time shifts has probability $0$ or $1$. For a Markov process this is equivalent to requiring that the only functions $f\in L^2(\pi)$ with $P_tf=f$ for all $t\ge0$ be $\pi$-almost surely constant, where $P_tf(x)=\E[f(X_t)\mid X_0=x]$ denotes the transition semigroup. A one-dimensional diffusion whose scale function diverges at both endpoints of its state space and whose speed measure is finite is ergodic with respect to its normalised speed measure \cite{Kutoyants2004}. The proof of \cref{thm:gamma} shows that both conditions hold for \eqref{eq:logistic} on $(0,\infty)$ when $\sigma^2<1$, and that the normalised speed measure is $\pi$. Moreover, since $\mathfrak m$ is a zero-flux stationary solution of the Fokker--Planck equation, $X$ is reversible with respect to $\pi$, so the generator $\mathcal L$ of $X$ is self-adjoint and non-positive on $L^2(\pi)$ \cite{BakryGentilLedoux2014}. Now let $f$ be bounded and not $\pi$-almost surely constant, put $\bar f=f-\int f\,\mathrm d\pi$, and consider the autocovariance
\[
  c_f(t)=\Cov\bigl(f(X_0),f(X_t)\bigr)=\langle\bar f,P_t\bar f\rangle_{L^2(\pi)} .
\]
By the spectral theorem, $c_f(t)=\int_{[0,\infty)}e^{-\varkappa t}\,\nu_f(\mathrm d\varkappa)$, where $\nu_f$ is the spectral measure of $\bar f$ with respect to $-\mathcal L$, a non-negative measure of total mass $\|\bar f\|^2_{L^2(\pi)}>0$. Its mass at $\varkappa=0$ is the squared norm of the orthogonal projection of $\bar f$ onto $\ker\mathcal L$. By ergodicity, $\ker\mathcal L$ consists of the constant functions, to which $\bar f$ is orthogonal, so $\nu_f(\{0\})=0$. Hence $c_f'(t)=-\int\varkappa e^{-\varkappa t}\,\nu_f(\mathrm d\varkappa)<0$, so $c_f$ is strictly decreasing on $[0,\infty)$. Since $c_f$ is known once $(K,\sigma)$ are, the value $c_f(rh)$, which is determined by the law of the pair, pins down $rh$, and hence $r$. For (iii), \cref{prop:clock} gives $\E[X^{(r)}_t]=\mathcal E(rt)$, where $\mathcal E$ is the mean curve at $r=1$, which depends only on $(K,\sigma)$ and the initial law. We first check that $\mathcal E(t)$ tends to the stationary mean $K(1-\sigma^2)$. The generator applied to $x^2$ gives $2r(1+\sigma^2)x^2-2rx^3/K$, which is bounded above by $C-x^2$ for some constant $C$. Hence $\sup_t\E[X_t^2]<\infty$, the family $(X_t)$ is uniformly integrable, and the convergence in law of a positive recurrent one-dimensional diffusion to its invariant law \cite{Kutoyants2004} carries over to the mean. Now, if two rates $r_1<r_2$ gave the same curve, then $\mathcal E(s)=\mathcal E(cs)$ for all $s$, with $c=r_2/r_1>1$. Iterating gives $\mathcal E(s)=\mathcal E(c^ks)$, which tends to the stationary mean as $k\to\infty$, so $\mathcal E$ would have to be constant.
\end{proof}

\Cref{thm:ident} assumes that the data are the exact values $X(t_k)$ of the process. In practice each measurement carries an error, and what is recorded is $Y_k=\log X(t_k)+\varepsilon_k$, with independent errors $\varepsilon_k$. \Cref{sec:regression} shows that part (ii) of the theorem does not survive this change: for the linearised dynamics, the joint law of two consecutive noisy measurements no longer determines the relaxation rate (\cref{prop:noise}).

\section{Observed-state estimation and its limitations under measurement noise}
\label{sec:regression}

\subsection{The observed-state lift is a regression}

The simplest way to fit \eqref{eq:glv} to a time series is to treat the measured log-abundances as the true states and to approximate each transition between consecutive samples by one Euler step. In the log-coordinates the Euler step is Gaussian, with mean $\zeta+h(c+Ae^{\zeta})$ and covariance $h\,\mathrm{diag}(\sigma^2)$, so the negative log-likelihood of the observed increments is a sum of squared prediction errors. Because the states are taken as data, this is the situation of \cref{cor:observed}, and the gradient is a plain sum of one-step terms. More is true: the root of that gradient can be written down explicitly, and it is the familiar regression.

\begin{definition}[Observed-state Euler cost]
\label{def:observed}
Let $\Delta\zeta_{n,i}$, $n=1,\dots,N$, $i=1,\dots,S$, be measured log-increments over spacings $h_n$, pooled over any number of trajectories, and let $X_n$ be the measured abundances at the start of each increment. The \emph{observed-state Euler cost} with a common noise intensity $\sigma_i\equiv\sigma_0$ is the negative log-likelihood of these increments when the measurements are taken as the true states and each increment is approximated by one Euler step of \eqref{eq:glv},
\[
  J(r,A,\sigma_0)
  =
  \sum_{n,i}
  \Bigl(
  \tfrac12\log(2\pi\sigma_0^2 h_n)
  +
  \frac{\bigl(\Delta\zeta_{n,i}-h_n(r_i+(AX_n)_i-\sigma_0^2/2)\bigr)^2}{2\sigma_0^2 h_n}
  \Bigr).
\]
An estimator obtained by minimising such a cost is called an \emph{observed-state} estimator.
\end{definition}

\begin{theorem}[The observed-state lift and the regression coincide]
\label{thm:ols}
In the setting of \cref{def:observed}, assume that the $N\times(S+1)$ matrix with rows $\sqrt{h_n}(1,X_n^\top)$ has full column rank and that the residuals below are not all zero. Let $(\widehat c,\widehat A)$ be the least-squares fit of the per-capita rates $\Delta\zeta_{n,i}/h_n$ on $(1,X_n)$ with weights $h_n$, and let $\mathrm{RSS}_w$ be its weighted residual sum of squares. Then $J$ has a single critical point, its global minimum, at
\[
  \widehat\sigma_0^2=\mathrm{RSS}_w/(NS),\qquad \widehat r_i=\widehat c_i+\tfrac12\widehat\sigma_0^2,
\]
where $N$ is the number of increments. For the logistic equation ($S=1$, with $\sigma_0^2$ playing the role of $s$) this point corresponds to admissible parameters if and only if $\widehat c>0$ and $\widehat A_{11}<0$, and it is then given by \eqref{eq:inverse}.
\end{theorem}

\begin{proof}
The key observation is that the model becomes linear once we absorb the It\^{o} correction into the intercept. Replacing $r_i$ by $c_i=r_i-\sigma_0^2/2$ is a smooth, invertible change of variables, so it neither creates nor removes critical points. In the new variables the predicted increment $h_n(c_i+(AX_n)_i)$ is linear in $(c,A)$, and $J$ is the negative log-likelihood of an ordinary Gaussian linear model. For any fixed $\sigma_0$, minimising over $(c,A)$ is a weighted least-squares problem. Under the rank assumption it has a unique solution, and that solution does not involve $\sigma_0$. Substituting it, the cost reduces to $\tfrac{NS}{2}\log\sigma_0^2+\mathrm{RSS}_w/(2\sigma_0^2)$, which has a single minimum at $\widehat\sigma_0^2=\mathrm{RSS}_w/(NS)$. Conversely, at any critical point of $J$ the derivative with respect to $(c,A)$ vanishes, which forces the least-squares solution, and the derivative with respect to $\sigma_0$ then forces $\widehat\sigma_0$. So there is exactly one critical point. For $S=1$ the relabelling \eqref{eq:inverse} maps positive $(a,b,s)$ one-to-one onto admissible $(r,K,\sigma)$, which gives the last statement.
\end{proof}

\begin{remark}[Bias of the noise estimate]
\label{rem:dof}
The value $\widehat\sigma_0^2$ in \cref{thm:ols} is the maximum-likelihood estimate and, as in any Gaussian linear model, it is biased downwards, because the $S(S+1)$ fitted coefficients $(\widehat c,\widehat A)$ absorb part of the noise. If the regressors $X_n$ were non-random and the Euler model exact, $\mathrm{RSS}_w/\sigma_0^2$ would follow a $\chi^2$ law with $S(N-S-1)$ degrees of freedom, since each of the $S$ species contributes $N$ residuals and $S+1$ fitted coefficients. Then $\E[\widehat\sigma_0^2]=\sigma_0^2\bigl(1-(S+1)/N\bigr)$, and the bias is removed by dividing by the number of degrees of freedom instead of the number of residuals,
\[
  \widetilde\sigma_0^2=\frac{\mathrm{RSS}_w}{NS-S(S+1)}=\frac{N}{N-S-1}\,\widehat\sigma_0^2,
  \qquad
  \widetilde r_i=\widehat c_i+\tfrac12\widetilde\sigma_0^2 .
\]
In the present setting the regressors are lagged states rather than fixed quantities, so $\widetilde\sigma_0^2$ is not exactly unbiased, but it removes the bias to leading order, as for autoregressions. The two estimates differ little when $N\gg S$.
\end{remark}

In our implementation the gradient of \cref{cor:observed} vanishes at the regression estimate to eleven digits, as it should. The theorem places the existing regression methods \cite{Stein2013,FisherMehta2014,Bucci2016} within the present framework. It also makes their statistical behaviour transparent, and Section~S1 of the Supplementary Material works it out in detail. With exact observations, the relative error of the estimated growth rate is $\sqrt{2\tau/T}$ for a record of length $T$, so what matters is how many relaxation times the record spans. With a finite sampling interval the estimates are biased: for the logistic equation $\widehat r$ converges to about $r[1-rh(1-\sigma^2)(\tfrac12+\sigma^2)]$. And on short records the usual small-sample bias of autoregressions appears. All of this concerns exact observations, however. We now turn to the real problem.

\subsection{Measurement noise}

Suppose the measured log-abundance is the true one plus independent noise, $Y_k=\zeta(t_k)+\varepsilon_k$ with $\varepsilon_k\sim\mathcal N(0,\eta^2)$. What changes in the analysis above? Qualitatively, the answer has been known in population ecology for a long time. Observation error biases estimates of density dependence and is hard to separate from process noise, which is why state-space models were introduced \cite{Dennis2006,Ives2003}. What we add here is an exact statement of how much can and cannot be learned from consecutive samples, and of where the regression ends up. The answer is clearest for the linearised dynamics. Close to its stationary state, the log-abundance of the logistic model behaves like an Ornstein--Uhlenbeck process, relaxing at rate $\kappa=r(1-\sigma^2)$ (\cref{sec:models}) with stationary variance $V\simeq\psi'(\alpha)$, the variance of $\log X$ under the Gamma law ($\psi'$ is the trigamma function). For this process everything can be computed exactly.

\begin{proposition}[Consecutive pairs under measurement noise]
\label{prop:noise}
Let $\zeta$ be a stationary Ornstein--Uhlenbeck process with relaxation rate $\kappa$, mean $\mu$ and variance $V$, observed at spacing $h$ as $Y_k=\zeta(kh)+\varepsilon_k$, with independent noise of variance $\eta^2$. Write $V_Y=V+\eta^2$ for the variance of a measurement, $\theta=V/V_Y\in(0,1]$ for the fraction of that variance which is signal, and $\rho_Y(h)=\theta e^{-\kappa h}$ for the correlation between consecutive measurements.
\begin{enumerate}
\item[(i)] The joint law of two consecutive measurements depends on the parameters only through $\mu$, $V_Y$ and $Ve^{-\kappa h}$. Consequently every relaxation rate $\kappa'\in(0,\kappa_{\max}]$, with $\kappa_{\max}=-h^{-1}\log\rho_Y(h)$, produces exactly the same pairs when combined with a suitable noise level: slower rates with more noise, faster ones with less. From pairs one can only conclude that the relaxation rate of the log-fluctuations satisfies $\kappa\le\kappa_{\max}$.
\item[(ii)] The one-step regression, that is, the least-squares fit of $Y_{k+1}-Y_k$ on $Y_k$, converges almost surely, as the number of samples grows, to the relaxation rate $\widehat\kappa=(1-\rho_Y(h))/h$ and the diffusion coefficient $\widehat s=V_Y(1-\rho_Y(h)^2)/h$. Without noise this underestimates $\kappa$, because $(1-e^{-\kappa h})/h<\kappa$. As the noise grows, $\widehat\kappa$ approaches $1/h$, a value set by the sampling interval rather than by the dynamics.
\item[(iii)] Covariances at two different non-zero lags, for instance the joint law of three consecutive measurements, identify all parameters:
\[
  e^{-\kappa h}=\rho_Y(2h)/\rho_Y(h),\qquad V=V_Y\,\rho_Y(h)\,e^{\kappa h},\qquad \eta^2=V_Y-V .
\]
The exact likelihood of the whole record is computed by the Kalman filter.
\end{enumerate}
\end{proposition}

\begin{proof}
Two consecutive measurements form a Gaussian pair with common mean $\mu$, common variance $V_Y$, and covariance $Ve^{-\kappa h}$. The noise adds to the variances but not to the covariance, because it is independent of everything else. A Gaussian pair is described completely by these numbers, which proves the first part of (i). Now take any rate $\kappa'>0$ and set the latent variance to $V'=Ve^{-(\kappa-\kappa')h}$, so that the covariance $V'e^{-\kappa'h}$ is unchanged, and the noise to $\eta'^2=V_Y-V'$, so that the total variance is unchanged. The pair law is then identical. The only constraint is that the new noise variance must be non-negative, i.e.\ $V'\le V_Y$, which is equivalent to $e^{-\kappa'h}\ge\rho_Y(h)$, i.e.\ $\kappa'\le\kappa_{\max}$. Slower dynamics are thus perfectly compensated by more noise, and faster dynamics, up to $\kappa_{\max}$, by less.

For (ii), the measurements form a stationary Gaussian ARMA$(1,1)$ sequence, which is ergodic, so sample moments converge almost surely to their expectations. The regression slope of $Y_{k+1}-Y_k$ on $Y_k$ therefore converges to $\Cov(Y_1-Y_0,Y_0)/V_Y=\rho_Y-1$, which corresponds to the rate $(1-\rho_Y)/h$. The residual variance converges to $\Var(Y_1-Y_0)-(1-\rho_Y)^2V_Y=V_Y(1-\rho_Y^2)$. The limiting cases follow by setting $\theta=1$ (no noise) and letting $\theta\to0$ (noise dominates).

For (iii), note again that the noise does not affect covariances at non-zero lag. Hence $\Cov(Y_0,Y_1)=Ve^{-\kappa h}$ and $\Cov(Y_0,Y_2)=Ve^{-2\kappa h}$, and their ratio isolates $e^{-\kappa h}$. The other two formulas follow. With irregular sampling the same argument applies to any two distinct lags $h_1\neq h_2$, whose covariance ratio is $e^{-\kappa(h_2-h_1)}$. The model is a linear Gaussian state-space model, whose likelihood is evaluated exactly by the Kalman filter \cite{Sarkka2013}.
\end{proof}

For the logistic equation the regression reports $\widehat r=\widehat a+\widehat s/2$ (\cref{thm:ols}). In the linearised picture its intercept plays the role of the relaxation rate, $\widehat a\approx\widehat\kappa$, because $\kappa=r(1-\sigma^2)=a$, so that $\widehat r\approx\widehat\kappa+\widehat s/2$ with $\widehat\kappa$ and $\widehat s$ given by (ii). The practical meaning is sobering. In the simulation design of \cref{sec:synth-log} the true relaxation time is $\tau=1/r=30$ hours. Part (ii) predicts that the regression should report $41.7$, $33.9$ and $24.9$ hours when the measurement noise has standard deviation $\eta=0$, $0.25$ and $0.5$ on the log scale, and the simulations give $43.7$, $35.0$ and $27.9$. For one particular noise level the regression happens to be right, but there is no way to know from the data whether one is at that level. With several species the noise does further damage. Errors in the regressors shrink the estimated interactions towards zero (the errors-in-variables effect), and the residual variance absorbs about $2\eta^2/h$, inflating the estimated environmental noise. The same reasoning applies to any estimator built from consecutive samples, not only to the regression. 
Part (iii) points to the way out. The information that separates dynamics from noise lies in correlations beyond consecutive pairs, and the natural way to use all of it at once is the likelihood of the whole record under a model for the hidden trajectory.

\section{The latent-state lift}
\label{sec:latent}

\subsection{Modelling the hidden trajectory}
\label{sec:latent-meaning}

We now take the measurement process into consideration and differentiate between observed and latent or hidden trajectories. The true log-abundances $\zeta(t)$ evolve according to \eqref{eq:glv} at all times. They are measured on the sampling days $t_0<t_1<t_2<\cdots$, which need not be equally spaced, and some taxa may be missing on some days. Each measurement is a noisy reading,
\begin{equation}
  Y_k=H_k\,\zeta(t_k)+\varepsilon_k,\qquad\varepsilon_k\sim\mathcal N(0,R_k),\qquad R_k=H_k\,\mathrm{diag}(\eta^2)\,H_k^\top,
  \label{eq:obs}
\end{equation}
where $\eta_i$ is the typical size of the measurement error of taxon $i$ on the log scale, and $H_k$ is the matrix that selects the taxa measured on day $k$ (the identity when all are measured). Known interventions are easy to include. An antibiotic course that reduces some taxa tenfold, for example, is a known shift of their log-abundances by $\log(0.1)$ at a known time. The unknowns are the dynamical parameters together with the noise levels, $u=(c,A,\sigma,\eta)$, which is $p=S+S^2+2S$ numbers when $\sigma$ and $\eta$ differ between taxa.

We collect these assumptions in a definition.

\begin{definition}[Latent-state model]
\label{def:latent}
The \emph{latent state} is the vector of true log-abundances $\zeta(t)$, a solution of \eqref{eq:glv}, possibly subject to known shifts $\zeta\mapsto\zeta+\Delta_q$ at known times. The \emph{measurements} are the $Y_k$ of \eqref{eq:obs}, with noises $\varepsilon_k$ independent of each other and of $\zeta$. Taxa not measured on day $k$ are simply absent from $Y_k$, which is why $Y_k$ involves the selection matrix $H_k$ and the measurement errors on that day have covariance $R_k$. An estimator that is a function of the likelihood of the measurements under this model is called a \emph{latent-state} estimator.
\end{definition}

It is important to keep three things apart here. There is the true trajectory $\zeta(t)$, which we never see. There are the measurements $Y_k$, which are what we have. And there is what we can \emph{infer} about the true trajectory from the measurements up to a given time. The regression of the previous section collapses the first two. The approach of this section keeps them distinct and works with the third.

That inference is carried out by a filter. At any moment the filter summarises what is known about $\zeta$ by a Gaussian belief, with mean $m$ (the best current guess of the log-abundances) and covariance $P$ (how uncertain that guess is). We call the pair $(m,P)$ the filter state. The numbers that the filter carries are not data but computed from the data and the model, and it therefore depends on the parameters. The filter updates its belief in three kinds of step that will be described below.

Between samples, the belief is carried forward by the model in small \emph{prediction steps} of length $\Delta t$. The mean follows the deterministic part of the dynamics. Because the log-abundance is uncertain, the abundance that enters the interaction term is averaged over the current belief, which for a Gaussian belief gives $\tilde x=\exp(m+\tfrac12\mathrm{diag}P)$. The uncertainty is transported by the linearised dynamics and grows by the environmental noise accumulated during the step. Writing $E=I+\Delta t\,A\,\mathrm{diag}(\tilde x)$ for the linearised one-step map and $\Pi(M)=\tfrac12(M+M^\top)$ for the symmetric part of a matrix,
\begin{equation}
  m^{+}=m+\Delta t\,(c+A\tilde x),\qquad
  P^{+}=\Pi\bigl(EPE^\top+\Delta t\,\mathrm{diag}(\sigma^2)\bigr).
  \label{eq:pred}
\end{equation}
The covariance update is the exact covariance of one Euler step of the linearised dynamics, which guarantees that $P$ stays positive definite. It is also exact to first order in $\Delta t$ for the non-linear Euler step: by Stein's lemma, $\Cov(\zeta,e^{\zeta_j})=P e_j\tilde x_j$ for a Gaussian $\zeta$ with covariance $P$, so the $\mathcal O(\Delta t)$ terms of $EPE^\top$ coincide with those of the covariance of $\zeta+\Delta t\,(c+Ae^{\zeta})$, and the approximation enters only at order $\Delta t^2$. Long gaps are simply bridged by more prediction steps.

When a sample arrives, the filter performs an \emph{update}. It compares its prediction $Hm$ (with $H=H_k$ the selection matrix of \cref{def:latent}) with the measurement, and the prediction error $v=Y_k-Hm$ is expected to have covariance $\Sigma=HPH^\top+R_k$: the filter's own uncertainty plus the measurement noise. The belief is then moved towards the measurement by an amount set by the Kalman gain $\mathcal K=PH^\top\Sigma^{-1}$, and the uncertainty shrinks accordingly:
\begin{equation}
  m^{+}=m+\mathcal K v,\qquad P^{+}=\Pi\bigl(P-\mathcal KHP\bigr),\qquad
  \ell_k=\tfrac12\log\det(2\pi\Sigma)+\tfrac12v^\top\Sigma^{-1}v .
  \label{eq:upd}
\end{equation}
The gain balances two sources of information. If the measurement noise is small compared with the filter's uncertainty, $\mathcal KH$ is close to the identity on the measured coordinates and the filter essentially adopts the measurement. If the measurement noise is large, $\mathcal K$ is small and the filter mostly keeps its own prediction. The quantity $\ell_k$ measures how surprised the filter is by the new measurement; it is the negative log-probability of $Y_k$ under the filter's prediction. Finally, at a known perturbation the filter applies a \emph{shift}, $m^{+}=m+\Delta_q$, leaving $P$ unchanged.

The cost we minimise is the total surprise, $J(u)=\sum_k\ell_k$. It is the negative log-likelihood of all measurements under the latent model. When there are many species we add a ridge penalty $\tfrac{\gamma_{\mathrm R}}{2}\sum_{i\ne j}A_{ij}^2$ on the interactions. The filter is approximate, because it replaces the true, non-Gaussian distribution of the hidden state by a Gaussian with the same mean and covariance (an assumed-density filter). It does, however, use every measurement in the context of all previous ones, which is exactly the information that \cref{prop:noise}(iii) shows is needed. For later reference we summarise the construction.

\begin{definition}[Filter and latent likelihood]
\label{def:filter}
A \emph{filter state} is a pair $Z=(m,P)$ of a vector $m\in\R^S$ and a symmetric positive definite matrix $P$, interpreted as the mean and covariance of a Gaussian belief about $\zeta$. Given $u$ and the measurements, the \emph{filter} is the sequence of states obtained from $Z_0=(Y_0,\mathrm{diag}(\widehat\Var\,Y))$ (missing entries of $Y_0$ replaced by sample means) by applying, in chronological order, prediction steps \eqref{eq:pred} across each gap (with $\lceil\text{gap}/\Delta t_{\max}\rceil$ equal sub-steps of length $\Delta t=\text{gap}/\lceil\text{gap}/\Delta t_{\max}\rceil\le\Delta t_{\max}$), an update \eqref{eq:upd} at each sampling day, and a shift $m\mapsto m+\Delta_q$ at each known perturbation. The \emph{latent likelihood cost} is $J(u)=\sum_k\ell_k$, the sum of the surprises of all updates. The first update uses $Y_0$ again. Its prediction error is zero, so its surprise $\tfrac12\log\det\bigl(2\pi(H_0P_0H_0^\top+R_0)\bigr)$ depends on $\eta$ only, and it acts as a weak data-dependent prior on the noise levels.
\end{definition}

\subsection{The gradient requires the backward sweep}

To maximise the likelihood we need its gradient, and here the structure of the filter matters. Consider how a parameter, say an interaction coefficient, influences the surprise at day $k$. It does so directly, through the last prediction steps before day $k$. It also acts indirectly: it influenced the belief on the previous day, which the filter has partly carried over, and so on back to the start of the record. Whether the influence survives each sample depends on how much the filter trusts the measurement, as the following proposition makes precise.

\begin{proposition}[Measurement noise gives the filter a memory]
\label{prop:memory}
In the update \eqref{eq:upd}, the sensitivity of the new mean to the old one is
\[
  \frac{\partial m^{+}}{\partial m}=I-\mathcal KH .
\]
If every taxon is measured ($H=I$) and the measurement noise vanishes ($\eta\to0$), this sensitivity tends to zero and the new covariance tends to zero: the filter discards its past at every sample. The likelihood then splits into independent pieces, one per interval between samples, and with a single prediction step per interval it is, up to a term that does not depend on $(c,A,\sigma)$, the observed-state Euler cost of \cref{def:observed} with species-specific noise intensities $\sigma_i$. If $\eta>0$, the sensitivity is not zero. For $H=I$ it equals $R(P+R)^{-1}$ with $R=\mathrm{diag}(\eta^2)$, a matrix with eigenvalues strictly between $0$ and $1$ (in one dimension, $\eta^2/(P+\eta^2)$). Part of the past belief is then carried through every sample.
\end{proposition}

\begin{proof}
From \eqref{eq:upd}, $m^+=m+\mathcal K(Y_k-Hm)$, so the derivative with respect to $m$ is $I-\mathcal KH$. (The gain depends on $P$, not on $m$.) With $H=I$ the gain is $\mathcal K=P(P+R)^{-1}$, hence $I-\mathcal K=R(P+R)^{-1}$. This matrix is similar to the symmetric matrix $R^{1/2}(P+R)^{-1}R^{1/2}$. When $R$ and $P$ are positive definite, $(P+R)^{-1}\prec R^{-1}$, so its eigenvalues lie in $(0,1)$. The predicted covariance is positive definite before every update, because each prediction step adds $\Delta t\,\mathrm{diag}(\sigma^2)\succ0$. Hence, when $R\to0$, the gain $P(P+R)^{-1}$ tends to the identity, so $m^+\to Y_k$ and $P^+=P-\mathcal KP\to0$, whatever the incoming belief. Every later surprise then depends on the parameters only through the steps taken after the most recent sample. Starting from $(m,P)=(Y_k,0)$, a single prediction step of length $h$ gives mean $Y_k+h(c+Ae^{Y_k})$, because $\tilde x=e^{Y_k}$ when $P=0$, and covariance $h\,\mathrm{diag}(\sigma^2)$. The next surprise is therefore the Gaussian Euler term of \cref{def:observed}, with species-specific $\sigma_i$ in place of the common $\sigma_0$. The only other term is the surprise of the first measurement, which involves the fixed initial belief and not $(c,A,\sigma)$.
\end{proof}

The measurement noise that makes a latent model statistically necessary (\cref{prop:noise}) is the same thing that gives the filter a memory, and it is this memory that ties the whole record together into one long chain of computed states. Without noise, the chain is cut at every sample, the cotangent lift reduces to the one-step formula of \cref{cor:observed}, and we recover the regression. With noise, sensitivities have to be propagated through every prediction step and every update, back to the start of the record. Long gaps add a second source of chaining, independent of the noise, because each gap is bridged by many prediction steps.

The likelihood is therefore exactly the kind of cost for which \cref{thm:lift} was designed. The steps are the prediction steps, updates and shifts, the state is the belief $(m,P)$, and each surprise $\ell_k$ is a cost attached to an update.

\begin{theorem}[Gradient of the latent likelihood]
\label{thm:latent}
Let $Z_i=(m_i,P_i)$ be the filter state before step $i$ (\cref{def:filter}), and let $\Phi_i$ be the corresponding prediction, update or shift. The latent negative log-likelihood is $J(u)=\sum_i\ell_i(Z_i,u)$, where $\ell_i$ is the surprise at update steps and zero otherwise, and the initial belief is fixed by the data. Its gradient is therefore given by the backward sweep of \cref{thm:lift}, which carries the covector $\lambda=(\bar m,\bar P)$ through the backward maps of each step listed in \cref{app:vjp}. One forward and one backward pass cost $\mathcal O(N_{\mathrm{steps}}S^3)$ operations, independently of the number of parameters. Finite differences would require $2p$ forward passes, that is $\mathcal O(N_{\mathrm{steps}}S^5)$ operations when the interaction matrix is estimated.
\end{theorem}

\begin{proof}
Each filter step is a smooth map of the belief and the parameters wherever $\Sigma$ is positive definite, and each surprise depends on the belief just before an update. This is the setting of \cref{thm:lift}, with no dependence of $\ell_i$ on the state after the step. One detail deserves comment: the symmetrisation $\Pi$ in \eqref{eq:pred}--\eqref{eq:upd}. Since $\Pi$ is an orthogonal projection, it is its own transpose, so in the backward sweep the covector attached to $P$ is symmetrised in exactly the same way. For the cost estimate, each backward map in \cref{app:vjp} involves a fixed number of products of $S\times S$ matrices and one inversion, which is $\mathcal O(S^3)$ work per step. Adding each step's contribution to the gradient costs $\mathcal O(S^2)$. A forward pass costs $\mathcal O(N_{\mathrm{steps}}S^3)$, and finite differences repeat it $2p$ times.
\end{proof}

Two clarifications help to place this result. First, reverse-mode automatic differentiation of the filter code produces the same numbers, because it performs the same backward sweep automatically \cite{SanzSerna2016,Kristensen2016}. We have checked this directly (\cref{sec:timing}): the two gradients agree to a relative difference of $4\times10^{-15}$. Writing the backward maps explicitly, as we do, makes the structure visible. It also forced us to treat the symmetrisation explicitly. An automatic tool differentiates $\Pi$ correctly by construction, but the instability of the unsymmetrised forward pass, discussed below, has to be addressed in any case. Second, there are gradient-free alternatives, but none is attractive here. Derivative-free optimisers do not scale to hundreds of parameters. The expectation--maximisation algorithm, which alternates a smoothing pass with a parameter update, has no closed-form parameter update for a non-linear drift, and it can converge slowly when the measurement noise is large, because the hidden states are then poorly determined by the data.

The complete estimation procedure is summarised in \cref{alg:latent}. Each iteration consists of one forward pass of the filter of \cref{def:filter}, which evaluates the cost, and one backward sweep, which returns its exact gradient (\cref{thm:latent}). The gradient is passed to a quasi-Newton optimiser. The iterations start from the regression estimate of \cref{thm:ols}, and they are restarted from several initial noise levels because the likelihood may have more than one local minimum in the noise parameters. At the optimum, the inverse of the Hessian, obtained by differencing the lifted gradient, provides approximate standard errors.

\begin{algorithm}[H]
\caption{Latent-state cotangent-lift estimator}
\label{alg:latent}
\begin{algorithmic}[1]
\Require samples $(t_k,Y_k)$, missing entries allowed; known shifts $(s_q,\Delta_q)$; maximal sub-step $\Delta t_{\max}$; ridge $\gamma_{\mathrm R}$
\Ensure $\widehat u$ minimising $J(u)+\tfrac{\gamma_{\mathrm R}}2\sum_{i\ne j}A_{ij}^2$, and its approximate covariance
\State Build the sequence of steps: prediction steps across each gap, an update at each sample, shifts at their times
\State Start from the regression of \cref{thm:ols} on one-day pairs; set $m_0=Y_0$, $P_0=\mathrm{diag}(\widehat\Var Y)$
\Repeat
\State Forward pass: run the filter, storing $(m_i,P_i)$ and accumulating $J$
\State Backward pass: apply the backward maps of \cref{app:vjp}, accumulating $\nabla_uJ$ (\cref{thm:latent})
\State Quasi-Newton (L-BFGS) step in $(c,A,\log\sigma,\log\eta)$
\Until{convergence; restart from several initial noise levels and keep the best}
\State Approximate covariance from the inverse Hessian, obtained by differencing the lifted gradient
\end{algorithmic}
\end{algorithm}

\subsection*{Implementation.} We wrote the forward and backward passes by hand and compiled them with Numba. Compared with finite differences, the lifted gradient agrees to better than $2\times10^{-8}$ in relative terms for up to forty species, including records with missing values, irregular gaps and perturbations. Three details turned out to be essential, and they are worth recording because they are easy to miss. The first concerns symmetry in the forward pass. In exact arithmetic $P$ stays symmetric, but the formulas \eqref{eq:pred}--\eqref{eq:upd} are not evaluated symmetrically in floating point. The round-off asymmetry then enters the inversion of $\Sigma$ and the gain, and in our runs with ten or more species the filter broke down after a few hundred samples unless $P$ was symmetrised after every step. The second is the backward counterpart of the first, and it is a matter of correctness rather than of round-off. Once $\Pi$ is part of the forward map, its cotangent lift, which is $\Pi$ again, must be applied to the covector of $P$. Omitting it gave gradients wrong by orders of magnitude. The third concerns positivity. The more common covariance step $P+\Delta t\,(FP+PF^\top)$, with $F=A\,\mathrm{diag}(\tilde x)$, can lose positive definiteness when the optimiser tries unusual parameters, whereas the form used in \eqref{eq:pred} cannot. When a trial point nevertheless produces an invalid $\Sigma$, we return a large cost together with the last valid gradient, which makes the line search step back.
\section{Numerical experiments and application to data}

\subsection{Simulated communities}
\label{sec:synthetic}

All simulated data in this section are generated from the continuous diffusions, integrated with very fine steps, and never from the filter or from the Euler scheme used by the regression. Otherwise the estimators would be tested on data produced by their own approximations.

\subsubsection{One taxon: recovering the relaxation time}
\label{sec:synth-log}

\begin{figure}[h!]
\centering
\includegraphics[width=\textwidth]{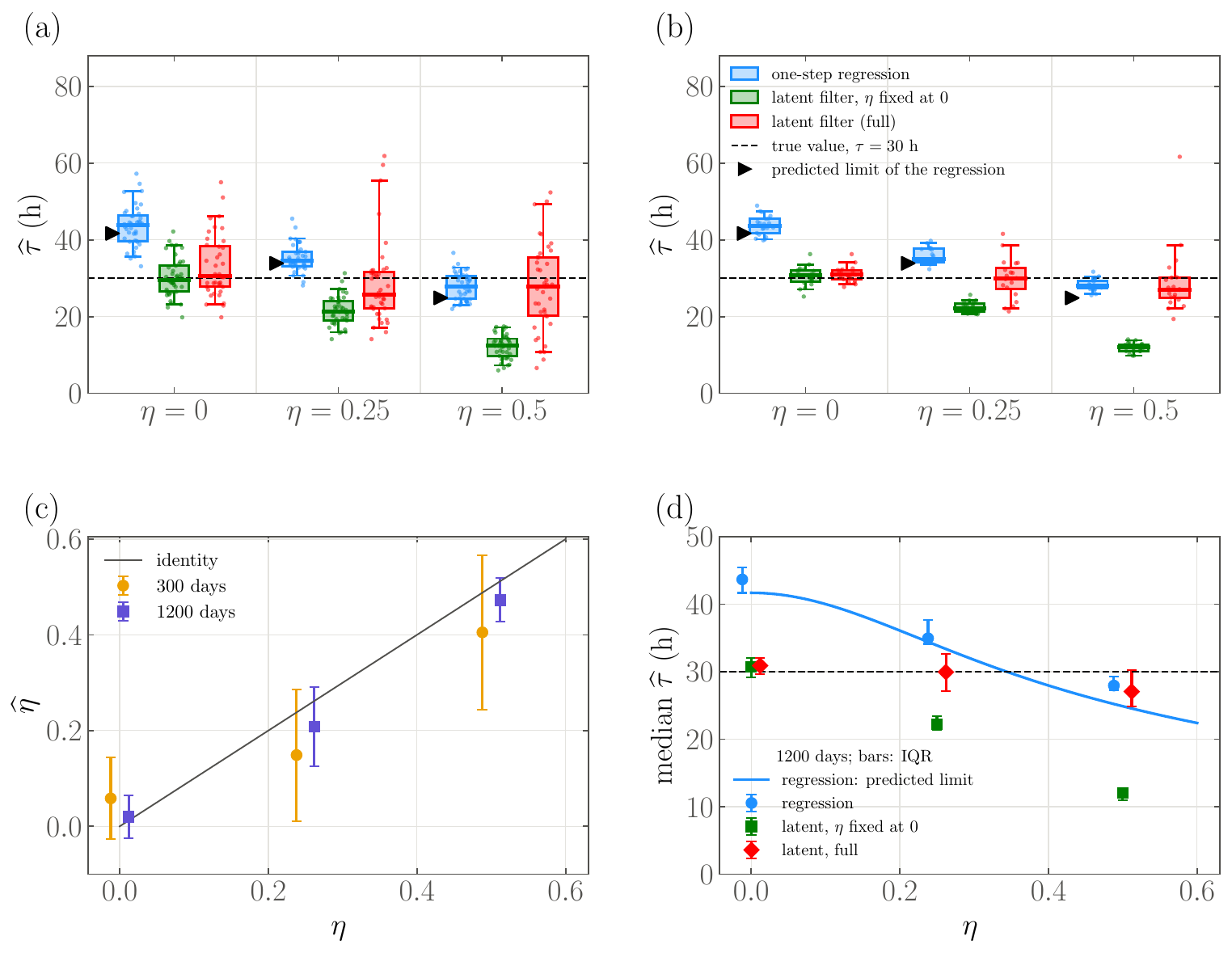}
\caption{A single taxon observed daily, with $20\%$ of the days missing and measurement noise of standard deviation $\eta$ on the log scale; the true relaxation time is $30$\,h (dashed lines). (a,b) Estimated relaxation times from the one-step regression, from the latent filter with fixed $\eta=0$ and from the full latent filter, for records of $300$ and $1200$ days. Dots are the individual replicates ($40$ and $20$ per setting). Black triangles mark the limit of the regression predicted by \cref{prop:noise}(ii). (c) Estimated against true measurement noise for the full latent filter. (d) Median estimates on the $1200$-day records (bars: interquartile range) as a function of the noise. The predicted limit of the regression is the continuous curve.}
\label{fig:latent-log}
\end{figure}

We start with the recovery of the relaxation time of a single taxon from noisy daily data. To this aim, we simulate the logistic equation with $\tau=30$ hours ($r=0.8$ per day), $K=0.05$ and $\sigma=0.45$, sample it daily for $300$ or $1200$ days, and delete $20\%$ of the days at random. We then add measurement noise of standard deviation $\eta=0$, $0.25$ or $0.5$ on the log scale. For comparison, the natural fluctuations of $\log X$ have standard deviation $0.54$, so the largest noise level is comparable to the signal, which is realistic for 16S data. Three estimators are compared. The first is the one-step regression of \cref{thm:ols}, using pairs of consecutive days. The second is the latent filter with the measurement noise forced to zero, which corrects for the coarse sampling but not for the noise. The third is the full latent filter. \Cref{tab:latent-log} and \cref{fig:latent-log} summarise $40$ replicates for the short records and $20$ for the long ones.

\begin{table}[h!]
\centering
\small
\setlength{\tabcolsep}{4pt}
\begin{tabular}{rr|c c|c|c|c}
\toprule
days & $\eta$ & regression & predicted limit & latent, $\eta\equiv0$ & latent & $\widehat\eta$ \\
\midrule
300 & 0 & $43.8\;[39.7,46.4]$ & $41.7$ & $29.6\;[26.5,33.4]$ & $30.5\;[27.8,38.3]$ & $0.06\pm0.08$\\
300 & 0.25 & $34.6\;[33.2,36.9]$ & $33.9$ & $21.2\;[18.9,24.1]$ & $25.7\;[22.1,31.6]$ & $0.15\pm0.14$\\
300 & 0.5 & $27.8\;[24.6,30.5]$ & $24.9$ & $12.4\;[9.7,14.2]$ & $27.8\;[20.2,35.4]$ & $0.41\pm0.16$\\
1200 & 0 & $43.7\;[41.7,45.5]$ & $41.7$ & $30.7\;[29.2,32.0]$ & $30.9\;[29.6,32.0]$ & $0.02\pm0.04$\\
1200 & 0.25 & $35.0\;[34.1,37.6]$ & $33.9$ & $22.2\;[21.3,23.4]$ & $29.9\;[27.1,32.6]$ & $0.21\pm0.08$\\
1200 & 0.5 & $27.9\;[27.3,29.3]$ & $24.9$ & $12.0\;[11.0,12.7]$ & $27.1\;[24.9,30.2]$ & $0.47\pm0.04$\\
\bottomrule
\end{tabular}
\caption{Estimated relaxation time in hours (median and interquartile range; the true value is $30$\,h) and estimated measurement noise (mean $\pm$ standard deviation). The predicted limit of the regression is \cref{prop:noise}(ii).}
\label{tab:latent-log}
\end{table}

The regression behaves exactly as \cref{prop:noise} predicts. Without measurement noise it overestimates the relaxation time (about $44$ hours), because a single Euler step over a whole day underestimates the speed of relaxation. As noise is added, its estimate drifts downwards, to $35$ and then $28$ hours, because day-to-day noise looks like fast dynamics. Longer records make the regression more precise but not more accurate: its estimate converges to the predicted limit, not to the truth. The second estimator, which corrects for the sampling but ignores the noise, is accurate in the absence of noise ($30.7$ hours). It fails badly as soon as noise is present, collapsing to about $12$ hours at $\eta=0.5$, because it has no choice but to attribute all variability to the dynamics. Only the full latent filter is accurate at every noise level: on the long records it gives $30.9$, $29.9$ and $27.1$ hours and recovers the noise level itself ($0.02$, $0.21$, $0.47$).

On the short records the picture is more nuanced, and it is important to say so. The latent estimate is roughly centred on the truth, but its spread is considerably larger than that of the regression. Telling slow dynamics from measurement noise requires the shape of the autocorrelation over several lags, and three hundred noisy days determine that shape only loosely. At the highest noise level and on short records, the regression even has the smaller overall error, simply because its bias happens to be small for that particular noise level. But nothing in the data tells the analyst that this is the case. Each fit takes between $0.3$ and $1.6$ seconds.

\subsubsection{Interacting communities}
\label{sec:synth-glv}

The second question is whether interactions can be recovered, and here the errors-in-variables effect enters. We draw communities of ten species from a random ensemble, $A=\gamma(-I+B)$ with off-diagonal entries $B_{ij}$ drawn independently from $\mathcal N(\mu/S,\sigma_A^2/S)$, using $\gamma=0.5$ per day, $\mu=-0.5$ and $\sigma_A=0.8$. To guarantee coexistence, we first choose the equilibrium abundances uniformly between $0.5$ and $1.5$ and set $r=-Ax^*$, and we discard draws whose equilibrium is unstable. The environmental noise is $\sigma_i=0.3$ and the measurement noise $\eta_i=0.25$ for every species. Sampling is daily for $300$ days, with $15\%$ of the days missing. To see the value of experimental perturbations, we also consider a design in which a random half of the species is reduced tenfold on days $60.5$, $120.5$, $180.5$ and $240.5$, between two sampling days, mimicking antibiotic pulses whose timing is known. The regression uses consecutive-day pairs that do not straddle a pulse. The latent filter uses all samples and treats the pulses as known shifts. Both use the same ridge penalty, $\gamma_{\mathrm R}=1$. The model, with a common environmental noise intensity and one measurement noise level per species, has $121$ parameters, and we repeat the experiment on $12$ independent communities per design.

\begin{table}[h!]
\centering
\small
\setlength{\tabcolsep}{4pt}
\begin{tabular}{ll|cccccc}
\toprule
design & method & slope & correlation & signs & self-interaction bias & $\widehat\sigma$ & $\widehat\eta$\\
\midrule
unperturbed & regression & $0.43\pm0.02$ & $0.57\pm0.02$ & $0.71$ & $-0.097\pm0.014$ & $0.371$ & -- \\
            & latent & $1.03\pm0.04$ & $0.62\pm0.03$ & $0.72$ & $-0.001\pm0.020$ & $0.255$ & $0.255$ \\
pulses      & regression & $0.49\pm0.02$ & $0.62\pm0.02$ & $0.74$ & $-0.087\pm0.010$ & $0.380$ & -- \\
            & latent & $1.01\pm0.04$ & $0.70\pm0.02$ & $0.75$ & $+0.043\pm0.010$ & $0.241$ & $0.269$ \\
\bottomrule
\end{tabular}
\caption{Ten-species communities with $\sigma=0.3$ and $\eta=0.25$ (mean $\pm$ standard error over $12$ communities). ``Slope'' is the regression slope of estimated on true off-diagonal interactions, so that $1$ means no systematic shrinkage. ``Correlation'' and ``signs'' measure agreement with the true interactions. ``Self-interaction bias'' is the mean error of the diagonal of $A$.}
\label{tab:latent-glv}
\end{table}

\begin{figure}[h!]
\centering
\includegraphics[width=\textwidth]{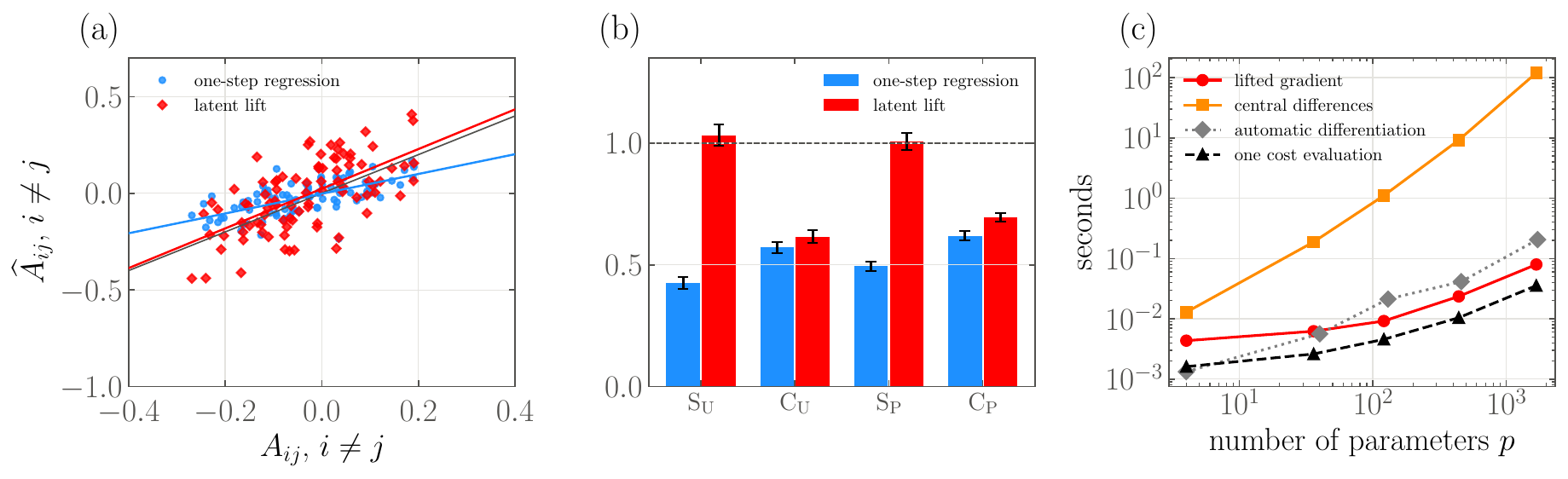}
\caption{Ten-species LV communities simulated with measurement noise.
The interaction matrices are $A=\tfrac12(-I+B)$, with off-diagonal entries $B_{ij}\sim\mathcal N(-0.5/S,\,0.64/S)$; the environmental noise is $\sigma_i=0.3$ and the measurement noise is $\eta_i=0.25$. Each community is sampled daily for $300$ days, with $15\%$ of the days missing. In the perturbed design, four known
pulses (days $60.5$, $120.5$, $180.5$ and $240.5$) each reduce a randomly chosen
half of the species tenfold. Both estimators use the ridge penalty
$\gamma_{\mathrm R}=1$.
(a)~Estimated against true off-diagonal interactions for one community with
pulses, with their least-squares lines; the thin line is the identity.
(b)~Mean $\pm$ standard error over $12$ communities of the attenuation slope (S),
that is, the slope of the least-squares line of estimated against true
off-diagonal interactions, and of the correlation (C) between estimated and true
off-diagonal interactions, without perturbations (U) and with pulses (P). The
dashed line marks the ideal value $1$.
(c)~Computing time of one gradient of the latent likelihood for $300$ daily
samples, at fixed parameter values, as a function of the number of parameters
$p$, for $S=1,5,10,20,40$ species: the hand-written cotangent lift (lifted
gradient), reverse-mode automatic differentiation of the same filter in JAX, and
central differences, whose time is estimated as $2p$ cost evaluations. The time
of one cost evaluation is shown for reference. The lifted-gradient and
automatic-differentiation timings come from separate runs; in a direct comparison
within one run the lifted gradient is $1.7$--$1.9$ times faster for $S\ge10$
(\cref{sec:timing}).}
\label{fig:latent-glv}
\end{figure}

The results (\cref{tab:latent-glv} and \cref{fig:latent-glv}a,b) show the errors-in-variables effect very clearly. The regression shrinks the interactions to less than half of their true size (slope $0.43$--$0.49$). It makes the self-limitation of each species look about $20\%$ stronger than it is (a mean error of $-0.10$ on diagonal entries of $-0.5$), which is the multi-species form of the drift towards fast relaxation in \cref{prop:noise}(ii): a measurement error that inflates $Y_k$ also deflates $Y_{k+1}-Y_k$. It also inflates the environmental noise from $0.30$ to about $0.37$, because it has to absorb the measurement noise somewhere. The latent filter removes the shrinkage almost exactly (slope $1.01$--$1.03$), removes the bias in the self-interactions in the unperturbed design and leaves a small bias of opposite sign ($+0.04$) with pulses, recovers the measurement noise, and improves the correlation with the true interactions and, marginally, the proportion of correctly identified signs. Perturbations help both methods, because pushing the community away from equilibrium makes the effect of each species on the others easier to see. They help the latent filter more, because it can use the samples around each pulse rather than discarding them. The price of removing the bias is a larger variance: the latent estimates scatter more around the truth (root-mean-square error $0.13$--$0.16$ against $0.10$ for the regression, whose shrinkage acts as a form of regularisation). The environmental noise is also underestimated by $15$--$20\%$, a residual effect of the Gaussian approximation in the filter and of the penalty.

It is equally important to know where the method stops working. With twenty species and the same amount of data the model has $441$ parameters, and the latent likelihood begins to over-fit. The fitted model explains the data better than the true parameters do, by more than chance alone would explain, and the estimated environmental noise collapses towards zero. The filter itself is not at fault. If we fix the interactions at their true values, it recovers $\sigma=0.283$ and $\eta=0.252$. The problem is that three hundred noisy daily samples simply do not contain enough information to determine four hundred interactions. Choosing the penalty by how well the model predicts the last quarter of the record ($\gamma_{\mathrm R}\approx100$--$300$) restores sensible behaviour, with a correlation of $0.55$--$0.58$ against $0.51$--$0.53$ for the regression, but both estimates are then strongly shrunk. At this scale, progress requires more information: sparsity, a prior built on the ensemble statistics of \cite{MegiasEtAl2026}, or more perturbation experiments.

\subsubsection{Computational cost of the gradient}
\label{sec:timing}

Finally, \cref{fig:latent-glv}(c) shows the computational advantage that motivates the cotangent lift. For $300$ daily samples we time one evaluation of the likelihood, one gradient by the backward sweep, and the $2p$ likelihood evaluations that finite differences would need. The backward sweep costs between $2.0$ and $2.7$ likelihood evaluations whatever the size of the community, from a single taxon ($4$ parameters) to forty species ($1681$ parameters). It is therefore $120$, $390$ and $1500$ times faster than finite differences for ten, twenty and forty species. A complete fit for ten species takes $6$ to $11$ seconds on a single core. With finite differences, a single gradient for forty species would take about two minutes.

The natural modern alternative is reverse-mode automatic differentiation, so we implemented the same filter in JAX and differentiated it automatically. The two gradients agree to a relative difference below $4\times10^{-15}$ for every community size, as they must, since automatic differentiation performs the same backward sweep. In speed the hand-written lift is $1.7$--$1.9$ times faster for ten to forty species ($0.12$\,s against $0.21$\,s per gradient for forty species), and it needs no compilation step (the automatic version takes $1.3$--$2.6$\,s to compile). For a single taxon the automatic version is faster, because its per-step overhead is smaller. The comparison confirms that the advantage of the backward sweep lies in its structure, not in how it is coded: both implementations cost a few likelihood evaluations per gradient, whatever the number of parameters.

\subsection{Application to gut microbiome time series}
\label{sec:data}

We now return to real data: the genus-level faecal time series of the two subjects (M3 and F4) of the Moving Pictures study \cite{Caporaso2011}, in the form distributed with \cite{Descheemaeker2020}. Counts are converted to relative abundances. For each subject we keep the six most abundant genera among those present in at least $85\%$ of the samples with a mean relative abundance of at least $2\%$. A first look at the sampling dates already shows why care is needed. The samples are not daily: for M3, $62$ of the $331$ intervals between samples are longer than a day, the longest being $17$ days, and for F4 the same holds for $16$ of $129$ intervals. The latent filter uses every sample and bridges the gaps with prediction steps of at most $\Delta t_{\max}=0.1$ days for the single-genus models and $0.2$ days for the six-genus model. Days on which a genus has zero reads are treated as missing measurements rather than replaced by an arbitrary small number, which would create artificial outliers on the log scale. The regression, for comparison, uses pairs of samples exactly one day apart. Since the data are relative abundances, all models should be understood as effective descriptions, as in \cite{Grilli2020}.

\subsubsection{One taxon at a time}

\begin{figure}[H]
\centering
\includegraphics[width=\textwidth]{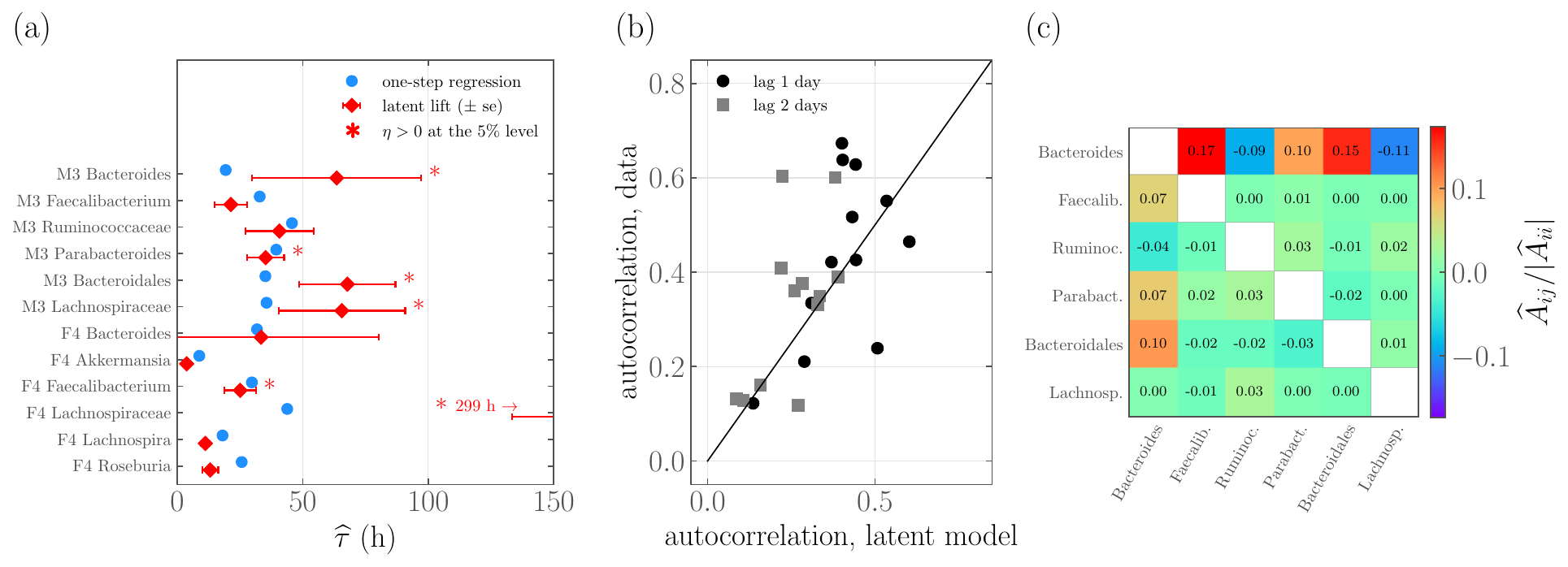}
\caption{Gut microbiome series of \cite{Caporaso2011} (subjects M3 and F4).
(a)~Relaxation time of each genus from the one-step regression and from the latent
lift ($\pm$ one standard error). An asterisk marks the series for which $\eta=0$
is rejected at the $5\%$ level; the estimate for F4 Lachnospiraceae ($299$\,h) is
off the scale.
(b)~Observed autocorrelation of the log-abundance at lags of one and two days
against the value implied by the fitted latent model, measurement noise included.
(c)~Interactions of the six-genus latent model for subject M3 ($\gamma_{\mathrm R}=3$,
fitted on the first $75\%$ of the record), shown as
$\widehat A_{ij}/|\widehat A_{ii}|$: the estimated effect of genus $j$ (column) on
genus $i$ (row), relative to its estimated self-limitation.}
\label{fig:latent-data}
\end{figure}

We first fit one latent logistic model to each genus (\cref{tab:data} and \cref{fig:latent-data}a,b). For half of the genera, six out of twelve, the data reject the hypothesis of negligible measurement noise. The most striking case is \emph{Bacteroides} in subject M3, the dominant genus of that community. The one-step regression assigns it a relaxation time of $19$ hours. The latent model attributes $79\%$ of the variance of its log-abundance to measurement noise and finds that the underlying population relaxes much more slowly, in about $64\pm34$ hours. The uncertainty is large, as it must be when most of the signal is noise, but even one standard error below the estimate the relaxation time exceeds the regression value. The autocorrelation of the measurements supports this reading. It is low, about $0.13$, and stays almost flat over one, two and three days. This is the signature of a slowly varying signal hidden under large uncorrelated noise, and it is reproduced by the latent model but by no noise-free logistic model with a relaxation time of about a day. Grilli obtained a relaxation time of about $19$ hours from daily microbiome series with a one-step estimator of this kind, regressing the next-day conditional mean on the current abundance \cite{Grilli2020}. Our analysis suggests that, for this genus at least, a value of that size reflects sequencing noise rather than ecological dynamics. Bacteroidales and Lachnospiraceae in M3 tell a similar story, with latent relaxation times of about $66$--$68$ hours, nearly twice the regression values. If the measurement noise is forced to zero, the filter reproduces the failure seen in the simulations: for every M3 genus it returns a shorter relaxation time than the regression, between $8$ and $31$ hours.

\begin{table}[H]
\centering
\scriptsize
\setlength{\tabcolsep}{2.6pt}
\begin{tabular}{llr|r|rr|rrr|cc|cc}
\toprule
 & & & regr. & \multicolumn{2}{c|}{latent} & & & & \multicolumn{2}{c|}{ACF lag 1} & \multicolumn{2}{c}{ACF lag 2}\\
Host & Taxon & $n$ & $\widehat\tau$ & $\widehat\tau$ & $\widehat\tau_{\eta\equiv0}$ & $\widehat\eta$ & noise & LR & data & model & data & model\\
\midrule
M3 & \emph{Bacteroides} & 332 & 19.3 & $63.6\pm33.7$ & 8.1 & $0.35\pm0.02$ & 0.79 & \textbf{6.5} & 0.12 & 0.14 & 0.13 & 0.11\\
M3 & \emph{Faecalibacterium} & 332 & 32.9 & $21.4\pm6.6$ & 18.7 & $0.19\pm0.18$ & 0.06 & 0.3 & 0.63 & 0.44 & 0.41 & 0.22\\
M3 & Ruminococcaceae & 332 & 45.7 & $40.7\pm13.6$ & 31.0 & $0.17\pm0.08$ & 0.14 & 1.0 & 0.55 & 0.54 & 0.35 & 0.33\\
M3 & \emph{Parabacteroides} & 330 & 39.4 & $35.2\pm7.4$ & 22.6 & $0.27\pm0.05$ & 0.11 & \textbf{4.9} & 0.52 & 0.43 & 0.36 & 0.26\\
M3 & Bacteroidales & 332 & 35.1 & $67.8\pm19.2$ & 18.0 & $0.38\pm0.03$ & 0.44 & \textbf{20.4} & 0.42 & 0.37 & 0.38 & 0.28\\
M3 & Lachnospiraceae & 332 & 35.6 & $65.6\pm25.3$ & 23.1 & $0.28\pm0.03$ & 0.40 & \textbf{5.8} & 0.43 & 0.44 & 0.33 & 0.33\\
F4 & \emph{Bacteroides} & 130 & 31.8 & $33.4\pm47.0$ & 20.0 & $0.10\pm0.09$ & 0.36 & 0.1 & 0.33 & 0.31 & 0.16 & 0.16\\
F4 & \emph{Akkermansia} & 130 & 8.8 & $3.7\pm0.8$ & 3.7 & $0.00\pm0.65$ & 0.00 & 0.0 & 0.46 & 0.60 & 0.39 & 0.39\\
F4 & \emph{Faecalibacterium} & 128 & 29.8 & $25.1\pm6.3$ & 18.0 & $0.34\pm0.09$ & 0.06 & \textbf{3.0} & 0.64 & 0.40 & 0.60 & 0.22\\
F4 & Lachnospiraceae & 130 & 43.8 & $299\pm165$ & 31.1 & $0.39\pm0.03$ & 0.40 & \textbf{15.6} & 0.67 & 0.40 & 0.60 & 0.38\\
F4 & \emph{Lachnospira} & 126 & 18.1 & $11.2\pm1.7$ & 11.2 & $0.00\pm0.42$ & 0.00 & 0.0 & 0.24 & 0.51 & 0.12 & 0.27\\
F4 & \emph{Roseburia} & 130 & 25.7 & $13.1\pm3.2$ & 13.1 & $0.00\pm2.99$ & 0.00 & 0.0 & 0.21 & 0.29 & 0.13 & 0.09\\
\bottomrule
\end{tabular}
\caption{Relaxation times in hours from one logistic model per taxon. $n$ is the number of samples with non-zero counts. ``Regr.'' is the one-step regression on one-day pairs; ``latent'' is \cref{alg:latent}, with standard errors from the inverse Hessian, and $\widehat\tau_{\eta\equiv0}$ is the same filter with the measurement noise set to zero. ``Noise'' is the share of the variance of $\log X$ attributed to measurement noise. LR is the likelihood-ratio statistic for $\eta=0$; values in bold are significant at the $5\%$ level (critical value $2.71$ for this boundary test). The last columns compare the autocorrelation function (ACF) of the measured $\log X$ at one and two days with that implied by the fitted latent model.}
\label{tab:data}
\end{table}

For the genera whose estimated measurement noise is exactly zero (\emph{Akkermansia}, \emph{Lachnospira} and \emph{Roseburia} in F4), the latent model differs from the regression only by its treatment of the sampling interval, and it finds shorter relaxation times ($4$--$13$ hours). This is the discretisation effect anticipated in \cref{prop:noise}(ii).

The model is not perfect, and the remaining discrepancies are informative. The autocorrelations at one and two days are reproduced for most genera (\cref{fig:latent-data}b). F4 Lachnospiraceae and both \emph{Faecalibacterium} series, however, are more persistent over two days than a single relaxation allows. For F4 Lachnospiraceae the latent relaxation time, about twelve days with a large uncertainty, points to a slow component that the logistic model captures only crudely. The measurement model is also deliberately simple: log-normal noise of constant size ignores the fact that sequencing data are counts. Each of these refinements, whether a count-based measurement model, a slowly varying environmental driver or a second time scale, would change the filter steps but not the method, and the gradient would still be given by the backward sweep of \cref{thm:latent}.

\subsubsection{Interactions within each subject}
\label{sec:data-glv}

Finally, we fit a six-genus latent Lotka--Volterra model to each subject, with separate environmental and measurement noise for each genus ($54$ parameters). To decide whether the interactions are supported by the data, we fit the model to the first three quarters of each record and ask how well it predicts the last quarter, one sample at a time. We compare a model without interactions (independent logistic equations) to one with a full interaction matrix and a ridge penalty of $3$ or $30$. For subject M3 the interactions help. The predictive surprise per held-out observation drops from $0.678$ to $0.653$ with the lighter penalty, about ten units of log-likelihood over the $384$ held-out measurements. For subject F4, whose record is shorter, they do not ($0.829$ without interactions against $0.847$ and $0.864$ with them). \Cref{fig:latent-data}(c) shows the M3 interaction matrix, with each row scaled by the self-limitation of the affected genus. The strongest couplings involve \emph{Bacteroides}, mostly as the genus affected: \emph{Faecalibacterium}, Bacteroidales and \emph{Parabacteroides} appear to favour it, and Lachnospiraceae and Ruminococcaceae to hinder it. \emph{Bacteroides} in turn appears to favour \emph{Faecalibacterium}, \emph{Parabacteroides} and the other Bacteroidales. Because the data are relative abundances, some of these couplings may reflect the constraint that proportions sum to one rather than genuine ecological interactions. We therefore read this result as evidence that a latent multi-species model predicts better for this subject, not as a measurement of who affects whom.

\section{Discussion}
\label{sec:discussion}

The cotangent lift is an exact and efficient way of differentiating a long computation with respect to many parameters. Applied to a cost built on measured states, it adds nothing beyond the chain rule, and for the logistic and Lotka--Volterra models its critical point is the regression already in use (\cref{thm:ols}). Its role appears when the cost involves states that must be inferred. Under measurement noise, consecutive pairs of observations do not determine the dynamics (\cref{prop:noise}), so no better one-step estimator can remove the bias; the latent likelihood does, and because measurement noise is exactly what gives the filter memory (\cref{prop:memory}), its gradient requires the backward sweep (\cref{thm:latent}).

It is worth being precise about what the latent model repairs and what it does not. It removes a bias, but it does not create information. In the limit of vanishing measurement noise it reduces to the regression (\cref{prop:memory}), and it therefore inherits the main weakness of any multiple regression with many coefficients, which is the variance of each individual estimate. Interactions can only be learned from how the abundances fluctuate around their equilibrium. To leading order each row of $A$ is estimated with covariance $\Sigma_1^{-1}/T$, where $\sigma^2\Sigma_1$ is the stationary covariance of the abundance fluctuations (Proposition~S1.2 of the Supplementary Material). Directions in which the community relaxes quickly fluctuate little and carry little information, and as the number of species grows this matrix tends to become ill-conditioned, much as correlated regressors do in any multiple regression. Each row has $S+1$ unknowns, and $S^2$ coefficients are assessed at once. The consequence is visible even in the most favourable setting we considered: with exact observations, $4000$ samples and ten species (Supplementary Section~S1.10), the standard deviation of a single interaction estimate is $0.32$, comparable to the true off-diagonal interactions, which lie between $-0.18$ and $-0.42$. Most of them would not be individually significant, although they are estimated without bias on average. Measurement noise makes matters worse, and the latent model makes them slightly worse still, because separating dynamics from noise is an additional task: in \cref{sec:synth-glv} removing the shrinkage raised the root-mean-square error of an interaction from $0.10$ to $0.13$--$0.16$. With twenty species the variance dominates and a penalty becomes unavoidable.

The practical conclusion is a division of labour. The latent model makes the largest difference for quantities described by a few parameters, such as relaxation times, noise levels and the share of the observed variability that is due to measurement. For these, the error of the regression is a bias that does not decrease with more data, whereas the latent estimate converges to the truth. Individual interactions, on the other hand, should be read with caution as soon as the community has more than a handful of species, and they should be reported together with their uncertainty, which the inverse Hessian provides at little cost. Questions about interactions are then better asked of the interaction matrix as a whole than coefficient by coefficient. This is why, in \cref{sec:data-glv}, we asked whether including interactions improves the prediction of held-out data, a single and well-posed test, rather than which entries of $A$ differ from zero.

Several mathematical questions remain open. First, we have proved identifiability from noisy observations only for the linearised dynamics (\cref{prop:noise}(iii)); whether the latent logistic and Lotka--Volterra models are identifiable from the joint law of the observations at several lags, and under which conditions on the sampling design, is not known. Second, the consistency and asymptotic normality of the estimator based on the assumed-density filter have not been established. The filter is exact to first order in the sub-step (\cref{sec:latent}), but the effect of the Gaussian closure accumulated over many steps on the limit of the estimator is not controlled, and the $15$--$20\%$ underestimation of $\sigma$ in \cref{sec:synth-glv} suggests that it is not negligible. Third, the same backward sweep applied to the gradient yields Hessian--vector products at a cost independent of $p$, and hence the Fisher information. This opens the way to choosing perturbations that are most informative about the interactions, a problem of optimal experimental design that attacks the variance limitation discussed above directly.

The approach has clear limitations. The filter approximates the distribution of the hidden state by a Gaussian. This approximation is accurate in our tests with a single taxon, but it leaves a residual bias in the environmental noise for ten species. Higher-order approximations or particle filters would reduce it, and the backward sweep applies to them as well. The measurement model ignores the count nature of sequencing data and the fact that relative abundances are constrained to sum to one. Including measured total biomass, or modelling log-ratios rather than log-abundances, would be the natural next step. The most fundamental limitation is the one just discussed. With a few hundred noisy daily samples, a community of twenty species has more interactions than the data can determine, and ecological prior knowledge has to enter the estimation. That knowledge could come from random-matrix theory \cite{MegiasEtAl2026}, from sparsity, or from deliberately designed perturbations. The data analysed here come from two subjects without interventions. Densely sampled perturbation experiments are the obvious next test.

\section{Conclusions}
\label{sec:concl}

For the stochastic logistic and Lotka--Volterra models, fitting the Euler likelihood to measured abundances as if they were the true ones amounts to the familiar regression of log-increments on abundances. Its behaviour can be worked out explicitly, and under measurement noise it is biased in a way that no method based on consecutive pairs of samples can avoid. Treating the true abundances as hidden, and computing the likelihood with a filter that follows them through the gaps between samples, removes that bias. Because the filter remembers its past when measurements are noisy, the gradient of this likelihood is exactly the backward sweep of the cotangent lift, at a cost independent of the number of parameters. On the Caporaso series this shows that the day-to-day variability of the dominant genus of one subject is mostly measurement noise. The underlying dynamics are estimated to be about three times slower than one-step estimates suggest, although with a large uncertainty, and interactions improve prediction for that subject.

\appendix
\section{Covector maps of the filter steps}
\label{app:vjp}

This appendix gives, for each filter step, the cotangent lift of \cref{def:lift}. The covector carried by the backward sweep is $\lambda=(\bar m,\bar P)$, with $\bar m$ its component on the mean and $\bar P$ its component on the covariance. We write the pullbacks $(\bar m^{+},\bar P^{+})\mapsto(\bar m,\bar P)$ and the parameter increments $D_u\Phi^*(\bar m^+,\bar P^+)$ for the three steps of \cref{sec:latent}. Bars denote covectors. Covectors of symmetric matrices are symmetric, and the pairing is $\langle\bar P,\delta P\rangle=\mathrm{tr}(\bar P\,\delta P)$. All formulas were checked against central differences (relative error $<2\times10^{-8}$).

\subsection*{Prediction sub-step \eqref{eq:pred}.} Let $\tilde x=\exp(m+\tfrac12\mathrm{diag}P)$ and $E=I+\Delta t\,F$ with $F=A\,\mathrm{diag}(\tilde x)$. Then
\begin{align*}
  \bar F&=2\Delta t\,\bar P^{+}EP, &
  \bar{\tilde x}&=\Delta t\,A^\top\bar m^{+}+\bigl(\textstyle\sum_iA_{ij}\bar F_{ij}\bigr)_j,\\
  \bar m&=\bar m^{+}+\bar{\tilde x}\odot\tilde x, &
  \bar P&=\Pi\bigl(E^\top\bar P^{+}E+\tfrac12\,\mathrm{diag}(\bar{\tilde x}\odot\tilde x)\bigr),
\end{align*}
and the parameter increments are
\[
  \bar c\mathrel{+}=\Delta t\,\bar m^{+},\qquad
  \bar A\mathrel{+}=\Delta t\,\bar m^{+}\tilde x^\top+\bar F\,\mathrm{diag}(\tilde x),\qquad
  \bar\sigma_j\mathrel{+}=2\Delta t\,\sigma_j\,\bar P^{+}_{jj}.
\]
\emph{Derivation.} $\delta(EPE^\top)=\delta E\,PE^\top+E\,\delta P\,E^\top+EP\,\delta E^\top$. Pairing with the symmetric $\bar P^+$ gives $\mathrm{tr}(E^\top\bar P^+E\,\delta P)+2\,\mathrm{tr}(PE^\top\bar P^+\delta E)$, and $\delta E=\Delta t\,\delta F$. The dependence of $F$ and of $m^+$ on $\tilde x$, and of $\tilde x$ on $(m,\mathrm{diag}P)$, gives the remaining terms by the chain rule.

\subsection*{Update \eqref{eq:upd}.} Let $\Sigma=HPH^\top+R$ with $R=H\,\mathrm{diag}(\eta^2)H^\top$, $G=\Sigma^{-1}$, $v=Y-Hm$, $\omega=Gv$, and the gain $\mathcal K=PH^\top G$. Define
\[
  w=\mathcal K^\top\bar m^{+},\qquad
  Q=\tfrac12\bigl(G-\omega\omega^\top\bigr)-\sym(w\omega^\top)+\mathcal K^\top\bar P^{+}\mathcal K,\qquad
  \bar v=\omega+w,
\]
where $Q$ is the covector of $\Sigma$, and $\overline{PH^\top}=\bar m^{+}\omega^\top-2\bar P^{+}PH^\top G$. Then
\[
  \bar m=\bar m^{+}-H^\top\bar v,\qquad
  \bar P=\Pi\bigl(\bar P^{+}+H^\top QH+\overline{PH^\top}H\bigr),\qquad
  \bar\eta_j\mathrel{+}=2\eta_j\,(H^\top QH)_{jj}.
\]
\emph{Derivation.} With $\delta\ell=\tfrac12\mathrm{tr}((G-\omega\omega^\top)\delta\Sigma)+\omega^\top\delta v$, $\delta m^{+}=\delta m+\delta(PH^\top)\omega+PH^\top(G\,\delta v-G\,\delta\Sigma\,\omega)$ and $\delta P^{+}=\delta P-\delta(PH^\top)GHP-PH^\top G\,\delta(HP)+PH^\top G\,\delta\Sigma\,GHP$, collect the coefficients of $\delta\Sigma$, $\delta v$ and $\delta(PH^\top)$, and use $\delta\Sigma=H\,\delta P\,H^\top+H\,\mathrm{diag}(2\eta\,\delta\eta)H^\top$ and $\delta v=-H\,\delta m$.

\subsection*{Shift.} The pullback is the identity, and there is no parameter increment.

\subsection*{The symmetrisation.} The map $\Pi(M)=\tfrac12(M+M^\top)$ is the orthogonal projection, for the trace pairing, onto symmetric matrices, so it is its own adjoint. Its cotangent lift is therefore $\Pi$ itself. This is why every formula for $\bar P$ above ends with $\Pi$: the backward recursion is then the exact gradient of the forward map as implemented, symmetrisation included. In the forward pass, $\Pi$ removes the antisymmetric part that round-off would otherwise introduce and feed into the inversion of $\Sigma$ (see the implementation notes in \cref{sec:latent}).

\bibliographystyle{unsrt}
\bibliography{references_revised}

\end{document}